%% file: paper.tex
\documentclass[sigconf]{acmart}

\AtBeginDocument{%
  }

\copyrightyear{2024}
\acmYear{2024}
\setcopyright{acmlicensed}\acmConference[CCS '24]{Proceedings of the 2024 ACM SIGSAC Conference on Computer and Communications Security}{October 14--18, 2024}{Salt Lake City, UT, USA}
\acmBooktitle{Proceedings of the 2024 ACM SIGSAC Conference on Computer and Communications Security (CCS '24), October 14--18, 2024, Salt Lake City, UT, USA}
\acmDOI{10.1145/3658644.3690254}
\acmISBN{979-8-4007-0636-3/24/10}

\usepackage{xspace}
\usepackage{xcolor}
\usepackage{paralist}
\usepackage{wrapfig}
\usepackage{multirow}
\usepackage{listings}
\usepackage{makecell}
\usepackage{boxedminipage}
\usepackage{booktabs}
\usepackage{subcaption}
\usepackage{tabularx}
\usepackage{url}
\usepackage{amsmath}

\usepackage{breakurl} 
\usepackage{boldline}
\usepackage{hyperref}
\usepackage[linesnumbered,ruled,vlined,noend]{algorithm2e}
\usepackage[titletoc]{appendix}
\usepackage{pifont}
\usepackage{balance}
\usepackage[htt]{hyphenat}
\usepackage{graphicx}
\usepackage{tikz}
\usepackage{enumitem}

\usepackage{listings}
\usepackage{caption}

\newcommand*\emptycirc[1][1ex]{\tikz\draw (0,0) circle (#1);} 

\newcommand*\fullcirc[1][1ex]{\tikz\fill (0,0) circle (#1);} 

\renewcommand{\paragraph}[1]{\vspace{0.05in}\noindent{\bf{#1}.}}

\newcommand{\sys}{{\sc BaseMirror}\xspace}

\AtEndPreamble{
  \usepackage{hyperref}
\hypersetup{
    colorlinks = true,
    linkcolor = blue,
    anchorcolor = blue,
    citecolor = blue,
    filecolor = blue,
    urlcolor = blue
}
}
\usepackage{balance}
\usepackage[all]{nowidow}
\begin{document}
\input{macro}

\pagestyle{plain}


\title{\sysname: Automatic Reverse Engineering of Baseband Commands from Android's Radio Interface Layer
}

\titlenote{This is the extended version of the CCS 2024 paper with the same title.}

\author{Wenqiang Li}
\email{li.14488@osu.edu}
\authornote{Both authors contributed equally to this research.}
\affiliation{%
  \institution{The Ohio State University}
  \country{USA}
}

\author{Haohuang Wen}
\email{wen.423@osu.edu}
\authornotemark[2]
\affiliation{%
  \institution{The Ohio State University}
  \country{USA}
}

\author{Zhiqiang Lin}
\email{zlin@cse.ohio-state.edu}
\affiliation{%
  \institution{The Ohio State University}
  \country{USA}
}


\input{section/0-abstract}




\maketitle


\input{section/1-intro}
\input{section/2-background}

\input{section/3-overview}

\input{section/4-system}

\input{section/5-eval}

\input{section/6-attack}
\input{section/7-discussion}
\input{section/8-related}
\input{section/9-conclusion}
\begin{acks}
We would like to thank the anonymous reviewers for their constructive feedback. This research was supported in part by ARO award W911NF2110081, DARPA award N6600120C4020, and NSF awards CNS-2112471 and ITE-2326882. Any opinions, findings, conclusions, or recommendations expressed are those of the authors and not necessarily of the ARO, DARPA and NSF.
\end{acks}


\bibliographystyle{ACM-Reference-Format}
\bibliography{paper}


\input{section/10-appendix}

\end{document}

%% file: macro.tex
\newcommand{\bheading}[1]{{\vspace{4pt}\noindent{\textbf{#1}}}}
\newcommand{\iheading}[1]{{\vspace{4pt}\noindent{\textit{#1}}}} 

\newcounter{note}[section]
\renewcommand{\thenote}{\thesection.\arabic{note}}

\newcommand{\zq}[1]{{\bf\textcolor{blue}{$\ll$ZQ: {\sf #1}$\gg$}}}

\newcommand{\ZQ}[1]{\refstepcounter{note}{\bf\textcolor{blue}{$\ll$ZQ~\thenote: {\sf #1}$\gg$}}}

\everypar{\looseness=-1 }
\newcommand\st[1]{\ding{#1}}

\newcommand{\etal}{\emph{et al.}\xspace}
\newcommand{\etc}{\emph{etc}\xspace}
\newcommand{\ie}{\emph{i.e.}\xspace}
\newcommand{\eg}{\emph{e.g.}\xspace}

\newcommand{\figurewidth}{\columnwidth}
\newcommand{\secref}[1]{\mbox{Sec.~\ref{#1}}\xspace}
\newcommand{\secrefs}[2]{\mbox{Sec.~\ref{#1}--\ref{#2}}\xspace}
\newcommand{\figref}[1]{\mbox{Fig.~\ref{#1}}}
\newcommand{\tabref}[1]{\mbox{Table~\ref{#1}}}
\newcommand{\appref}[1]{\mbox{Appendix~\ref{#1}}}
\newcommand{\ignore}[1]{}

\newcommand{\gbytes}{\ensuremath{\mathrm{GB}}\xspace}
\newcommand{\mbytes}{\ensuremath{\mathrm{MB}}\xspace}
\newcommand{\kbytes}{\ensuremath{\mathrm{KB}}\xspace}
\newcommand{\bytes}{\ensuremath{\mathrm{B}}\xspace}
\newcommand{\hertz}{\ensuremath{\mathrm{Hz}}\xspace}
\newcommand{\ghertz}{\ensuremath{\mathrm{GHz}}\xspace}
\newcommand{\msecs}{\ensuremath{\mathrm{ms}}\xspace}
\newcommand{\usecs}{\ensuremath{\mathrm{\mu{}s}}\xspace}
\newcommand{\nsecs}{\ensuremath{\mathrm{ns}}\xspace}
\newcommand{\secs}{\ensuremath{\mathrm{s}}\xspace}
\newcommand{\gbits}{\ensuremath{\mathrm{Gb}}\xspace}

\newcommand{\allril}{28\xspace}
\newcommand{\allcmd}{TBD\xspace}
\newcommand{\uniquecmd}{873\xspace}
\newcommand{\allattack}{7\xspace}
\newcommand{\totalattack}{8\xspace}
\newcommand{\allevaluatecmd}{179\xspace}

\newcounter{packednmbr}
\newenvironment{packedenumerate}{
\begin{list}{\thepackednmbr.}{\usecounter{packednmbr}
\setlength{\itemsep}{0pt}
\addtolength{\labelwidth}{4pt}
\setlength{\leftmargin}{12pt}
\setlength{\listparindent}{\parindent}
\setlength{\parsep}{3pt}
\setlength{\topsep}{3pt}}}{\end{list}}

\newenvironment{packeditemize}{
\begin{list}{$\bullet$}{
\setlength{\labelwidth}{8pt}
\setlength{\itemsep}{0pt}
\setlength{\leftmargin}{\labelwidth}
\addtolength{\leftmargin}{\labelsep}
\setlength{\parindent}{0pt}
\setlength{\listparindent}{\parindent}
\setlength{\parsep}{0pt}
\setlength{\topsep}{3pt}}}{\end{list}}

\newcommand{\sysname}{\textsc{BaseMirror}\xspace}

\newcommand{\lf}{\textit{lockout factors}\xspace}

\newcommand{\threshold}{\ensuremath{\mathit{thrsh}}\xspace}

\newcommand{\cmark}{\ding{51}}%
\newcommand{\xmark}{\ding{55}}%

%% file: section/0-abstract.tex
\begin{abstract}

In modern mobile devices, baseband is an integral component running on top of cellular processors to handle crucial radio communications. However, recent research reveals significant vulnerabilities in these basebands, posing serious security risks like remote code execution. Yet, effectively scrutinizing basebands remains a daunting task, as they run closed-source and proprietary software on vendor-specific chipsets. Existing analysis methods are limited by their dependence on manual processes and heuristic approaches, reducing their scalability. 
This paper introduces a novel approach to unveil security issues in basebands from a unique perspective: to uncover vendor-specific baseband commands from the {\it Radio Interface Layer (RIL)}, a hardware abstraction layer interfacing with basebands. To demonstrate this concept, we have designed and developed \sys, a static binary analysis tool to automatically reverse engineer baseband commands from vendor-specific RIL binaries. It utilizes a bidirectional taint analysis algorithm to adeptly identify baseband commands from an enhanced control flow graph enriched with reconstructed virtual function calls. Our methodology has been applied to \allril vendor RIL libraries, encompassing a wide range of Samsung Exynos smartphone models on the market. Remarkably, \sys has uncovered \uniquecmd unique baseband commands undisclosed to the public. Based on these results, we develop an automated attack discovery framework to successfully derive and validate \totalattack zero-day vulnerabilities that trigger denial of cellular service and arbitrary file access on a Samsung Galaxy A53 device. These findings have been reported and confirmed by Samsung and a bug bounty was awarded to us.


\end{abstract}

%% file: section/1-intro.tex
\section{Introduction}
\label{sec:intro}

Baseband is a proprietary and mandatory component in mobile devices, responsible for overseeing all radio functions like voice calls, text messages (SMS), and cellular data connections. Essentially, the baseband serves as the primary interface enabling a smartphone to communicate with the external cellular network, specifically the base station towers. Consequently, the baseband firmware code is complicated, dealing with low-level signals and cellular states, which are transparent to end-users. To date, key manufacturers like Qualcomm, Samsung, and MediaTek have created closed-source software and integrated system-on-chips (SoCs) that are used by manufacturers such as Google and Apple to produce smartphones, tablets, and other cellular devices.

However, it is less known that the baseband operates as a self-contained and highly independent system with numerous undisclosed functionalities. Specifically, the baseband firmware runs on a dedicated {\it Cellular Processor (CP)} with a real-time operating system (RTOS), rendering it entirely separated from the mobile device's main processor, commonly referred to as the {\it Application Processor (AP)}~\cite{kim2021basespec}. While previous studies have identified various vendor-specific AT commands in smartphone firmware triggering unpublicized baseband functions~\cite{tian2018attention, karim2019opening}, these AT commands are typically used in legacy Android phone models. In fact, we notice that most recent Android devices use proprietary inter-process communication (IPC) mechanisms to communicate with the baseband~\cite{samsung_ipc,qualcomm_ipc}. These baseband interfaces also have raised significant security concerns. For example, Samsung Galaxy devices' baseband was reported to contain remote file access (RFS) backdoors, providing unauthorized access to read, write, or delete files on the smartphone's local file 
system~\cite{samsung_backdoor}.

Even more concerning, recent demonstrations have revealed that mobile basebands are susceptible to exploitation, allowing the triggering of vulnerabilities like remote code execution (RCE) over the air~\cite{grassi2018exploitation}, even extending to the latest 5G mobile devices~\cite{grassi2021over}. Over the years, it has become evident that the baseband often suffers from inadequate security engineering, showcasing memory corruption vulnerabilities~\cite{zeng2012design, golde2016breaking, hernandez_firmwire_2022, maier2020basesafe} and non-compliance with cellular specifications~\cite{kim2021basespec, hussain2021noncompliance}. Simultaneously, high-privileged malware on a device can inject malicious requests into the baseband, activating security-sensitive functions. These exploitation pathways pose a serious risk, allowing malicious actors to compromise the device and jeopardize the user's security and privacy.


Current approaches involve scrutinizing baseband firmware implementations through reverse engineering (RE) to identify and address undesired behaviors and vulnerabilities in basebands. As of now, only a few tools have been developed for baseband RE, encompassing both static analysis~\cite{kim2021basespec,liu2024semantic} and dynamic analysis~\cite{hussain2021noncompliance}, with some utilizing emulation methods~\cite{hernandez_firmwire_2022,maier2020basesafe}. However, these tools come with several limitations, often relying on manual analysis and heuristics, and thus do not generalize well~\cite{kim2021basespec,maier2020basesafe}. Indeed, baseband RE is an exceedingly difficult task due to various factors such as undisclosed implementation details, diverse architectures, and the absence of debugging symbols.

This paper presents a novel approach, \sys, which unveils baseband security issues from a new perspective. Instead of directly analyzing the baseband firmware, \sys reverse-engineers vendor-specific baseband commands from the \textit{Radio Interface Layer} (RIL)~\cite{android_ril}, which effectively serves as a \textit{mirror} of the baseband interfaces. This ``mirror-based'' perspective has been successfully applied in previous research to uncover security vulnerabilities in automotive and IoT devices~\cite{wen2020automated,wen2020plug,wen2020firmxray,wang2019looking}.
In this work, our key insight is rooted in the generic architecture of Android mobile devices, where all communication between the Application Processor (AP) and Cellular Processor (CP) is processed and mediated by the RIL—a hardware abstraction layer between the radio hardware and the operating system. The critical commands implemented in the RIL thus expose exploitable attack surfaces targeting the AP and CP.
As the RIL logic is typically integrated into a standalone shared library by the baseband hardware vendor (e.g., Samsung and Qualcomm), we consequently formulate it as a reverse engineering task to uncover vendor-specific baseband commands from the RIL binary code.



Nonetheless, developing such a static analysis pipeline poses three technical challenges. First, it requires a generic algorithm to comprehensively identify desired program paths generating these baseband commands so that they can be applied to various baseband versions and models without human intervention. Second, the inter-procedural call graph of the RIL binary needs to be recovered but it is challenging due to the massive use of virtual function calls~\cite{erinfolami2020devil,bacon1996fast} that could not be resolved by off-the-shelf reverse engineering tools by default~\cite{ghidra,ida,shoshitaishvili2016sok}. Third, it needs to accurately filter out commands not dedicated to the baseband, as the RIL binaries are still relatively complex, typically involving over 15K functions, and not all contribute to baseband communication.

We have addressed the above challenges and implemented a prototype of \sys as an automatic static binary code analysis tool to reveal vendor-proprietary baseband commands from the Android RIL. It employs a bidirectional taint analysis algorithm to comprehensively track the data flow from fundamental system APIs invoked for baseband interactions. To facilitate this analysis, \sys recovers the indirect function calls (especially virtual calls) by employing a lightweight and efficient algorithm to resolve the virtual call target and function. Subsequently, it discards commands not intended for the baseband, based on the system path associated with the corresponding command channels.

To evaluate \sys, we have tested it with \allril vendor RIL libraries extracted from real Android firmware images, covering a wide range of Samsung smartphone models with the proprietary Exynos baseband. \sys successfully extracted \uniquecmd unique vendor-specific commands. We further developed a dynamic validation framework on a real device to verify the correctness of \allevaluatecmd commands that do not involve external parameters, which shows no false positive. 
%
To assess the security implications of our results, we developed an automatic attack payload discovery tool that derives \totalattack new exploitable commands on a Samsung Galaxy A53. We demonstrate that these attack commands can be leveraged to trigger service disruption and denial-of-service attacks on the CP, including three new attacks that require re-flashing the baseband to recover. We also discover a new vulnerability enabling arbitrary access to the AP's file system. This vulnerability has been patched in the latest RIL firmware and a bug bounty was awarded.

%

In summary, we make the following contributions:

\begin{itemize}

    \item We present a new approach to unveil security issues in basebands from the Radio Interface Layer, by reverse engineering vendor-specific baseband commands.
    
    \item We developed \sys, a static binary analysis tool, and used that to extract \uniquecmd unique baseband commands from \allril different Samsung Exynos smartphones' firmware.

    \item We identified \totalattack zero-day vulnerabilities that exploit baseband commands to trigger denial-of-service attacks on the CP and arbitrary file accesses on the AP.  
        
    \item We have released a proof-of-concept implementation of \\ \sys at \url{https://github.com/OSUSecLab/BaseMirror}. 

\end{itemize}

%% file: section/2-background.tex


\section{Background and Motivation}
\label{sec:background_ril}

\begin{figure}[t]
    \centering
    \includegraphics[width=0.45\textwidth]{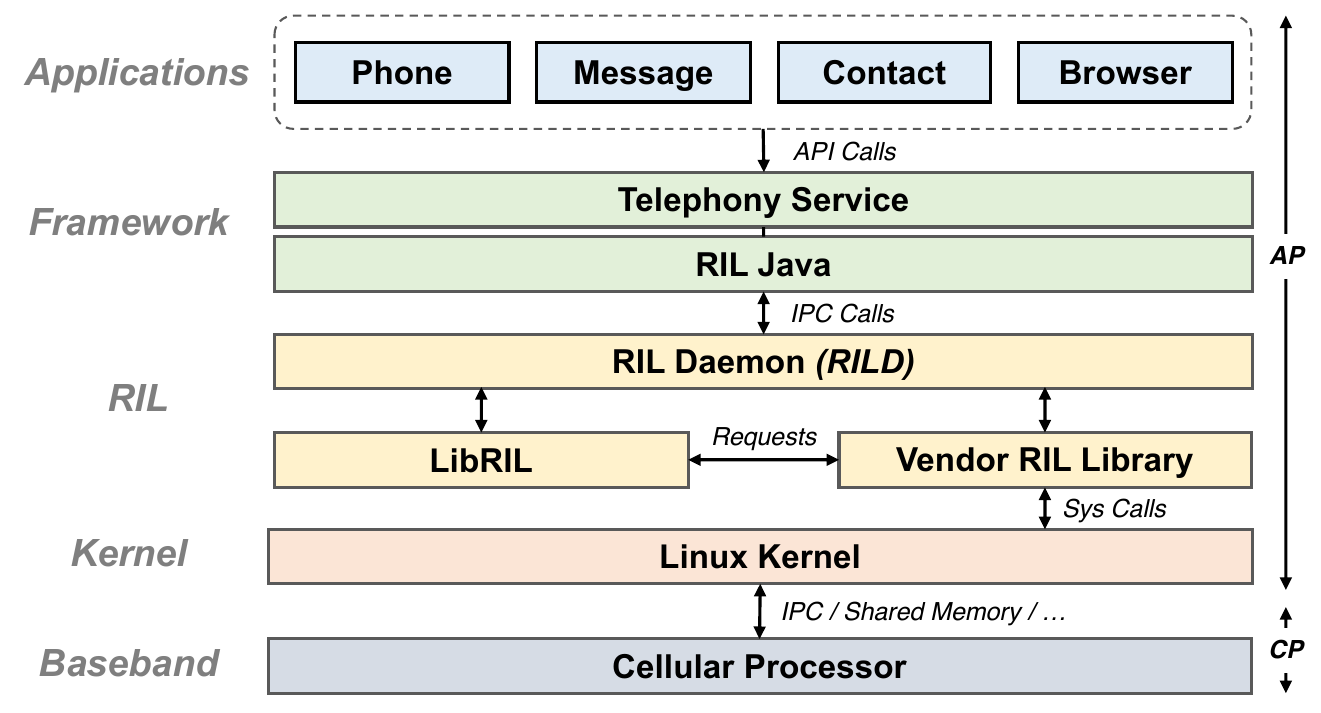}
    \caption{RIL architecture within an Android mobile device.}
    \label{fig:ril_arch}
\end{figure}

\noindent \textbf{Radio Interface Layer.} In modern wireless systems, the Radio Interface Layer (RIL), a crucial element of the Android architecture \cite{android_ril}, serves as the bridge between software and radio hardware. Initially specific to Android \cite{android_ril}, RIL now broadly represents similar concepts in the hardware abstraction layer (HAL) \cite{kroll2021aristoteles}. As a vital interface, RIL enables seamless communication between the operating system and the radio modem, or baseband, essential for data transmission and reception over wireless networks. By abstracting radio hardware complexities, RIL provides a standardized interface that allows higher-level software, especially telephony applications, to interact with the radio module effectively. Its importance extends beyond basic functionality, enhancing interoperability, streamlining communication protocols, and optimizing resource use in mobile devices. For a comprehensive understanding, we outline the RIL architecture within a typical Android system in \autoref{fig:ril_arch} and elaborate on the primary RIL components subsequently.

\paragraph{RILD}
The RIL Daemon (RILD) serves as the intermediary between the higher-level Android framework layer and the remaining RIL components~\cite{android_ril}. More specifically, the RILD initializes the vendor RIL and LibRIL during system start-up. The initialization will bridge the communication between vendor RIL and LibRIL as the RILD will register the vendor RIL's communication interfaces (e.g., a \texttt{RIL\_RadioFunctions} structure) to the LibRIL so that they can be invoked in the future. After the initialization, the RILD processes all communication events from the Android telephony service (via the RIL Java, or RILJ layer through Android's Inter-process communication mechanisms~\cite{android_ipc}) and further dispatches requests.

\paragraph{LibRIL}
The LibRIL is a generic Android shared library (\texttt{.so}) that interacts with the vendor RIL through standard interfaces defined in the Android RIL source code~\cite{android_ril}. To monitor the events and responses, LibRIL maintains an event loop thread that constantly listens to a series of file descriptors of interest. Once the LibRIL is notified of events from either the RILJ or the baseband, it triggers the corresponding handler functions to respond to the events. For instance, LibRIL will invoke the vendor RIL through its \texttt{onRequest} interface after receiving events from the upper framework layer. The Android Open Source Project (AOSP)~\cite{aosp} has defined a set of standard RIL commands to ensure basic cellular functionality, such as dialing and SMS. 


\paragraph{Vendor RIL and the Reverse Engineering Opportunities}
The Android platform, while maintaining generic components like RILD and LibRIL for standard RIL communication, also enables hardware vendors to craft their customized logic and protocols within a shared vendor RIL library \cite{android_ril}. This flexibility is vital due to the diversity of baseband chipsets used by different manufacturers, necessitating a vendor RIL that is adeptly tailored to the unique characteristics and functionalities of each chipset. These vendor-specific implementations are integrated into a shared library ({\tt .so}). For instance, AOSP provides a reference vendor RIL implementation utilizing the Hayes AT command set~\cite{aosp, tian2018attention}. Notably, while vendor RILs are permitted to employ various protocols for baseband communication, they must adhere to the standardized RIL interfaces to ensure compatibility with the generic RIL layer.

In this paper, our primary focus is on these vendor RILs, which present a unique and standard interface, offering a strategic vantage point for reverse engineering the proprietary baseband interfaces. By design, these vendor RIL libraries encompass many undisclosed commands and protocols with the baseband, which are not only critical in understanding the complicated workings of the baseband but also helpful in identifying potential security and privacy vulnerabilities inherent in these systems. Therefore, by dissecting and analyzing the vendor RIL, we can unveil hidden aspects of the baseband, thereby contributing to the broader understanding of its security posture and operational dynamics.


%% file: section/3-overview.tex
\section{Overview}
\label{sec:overview}

\subsection{Objectives and Assumptions}
\label{sec:assumption}

\noindent\textbf{Objectives.}
Our objective is to reverse engineer the \textit{vendor implemented RIL libraries} to identify {proprietary baseband commands used for communication between AP and CP}. To this end, we define two types of such command types based on the direction: (1) \textit{Solicited commands} issued by the AP to instruct the CP for dedicated functions and (2) \textit{Unsolicited commands} used by the CP to proactively interact with the AP via notification mechanisms. These proprietary commands are fundamentally exchanged via standard communication interfaces at the RIL. Essentially, our goal is to comprehensively identify these proprietary command payloads and demonstrate their security impacts. 

\paragraph{Assumptions}
Considering the availability of RIL firmware and open architecture, this paper focuses on Android RIL~\cite{android_ril}. We assume the smartphone vendors have followed the official RIL architecture, which allows us to identify and extract the proprietary vendor RIL libraries from the phone's factory images. We assume our targeted vendor RIL libraries can be disassembled and decompiled, and they are stripped and not obfuscated (which is true at this time of writing). As RIL resides at the native layer of Android, the associated libraries are developed with C++ language for ARM architecture that is widely adopted by most commodity smartphones.

\begin{figure}[t]
    \centering
    \includegraphics[width=0.44\textwidth]{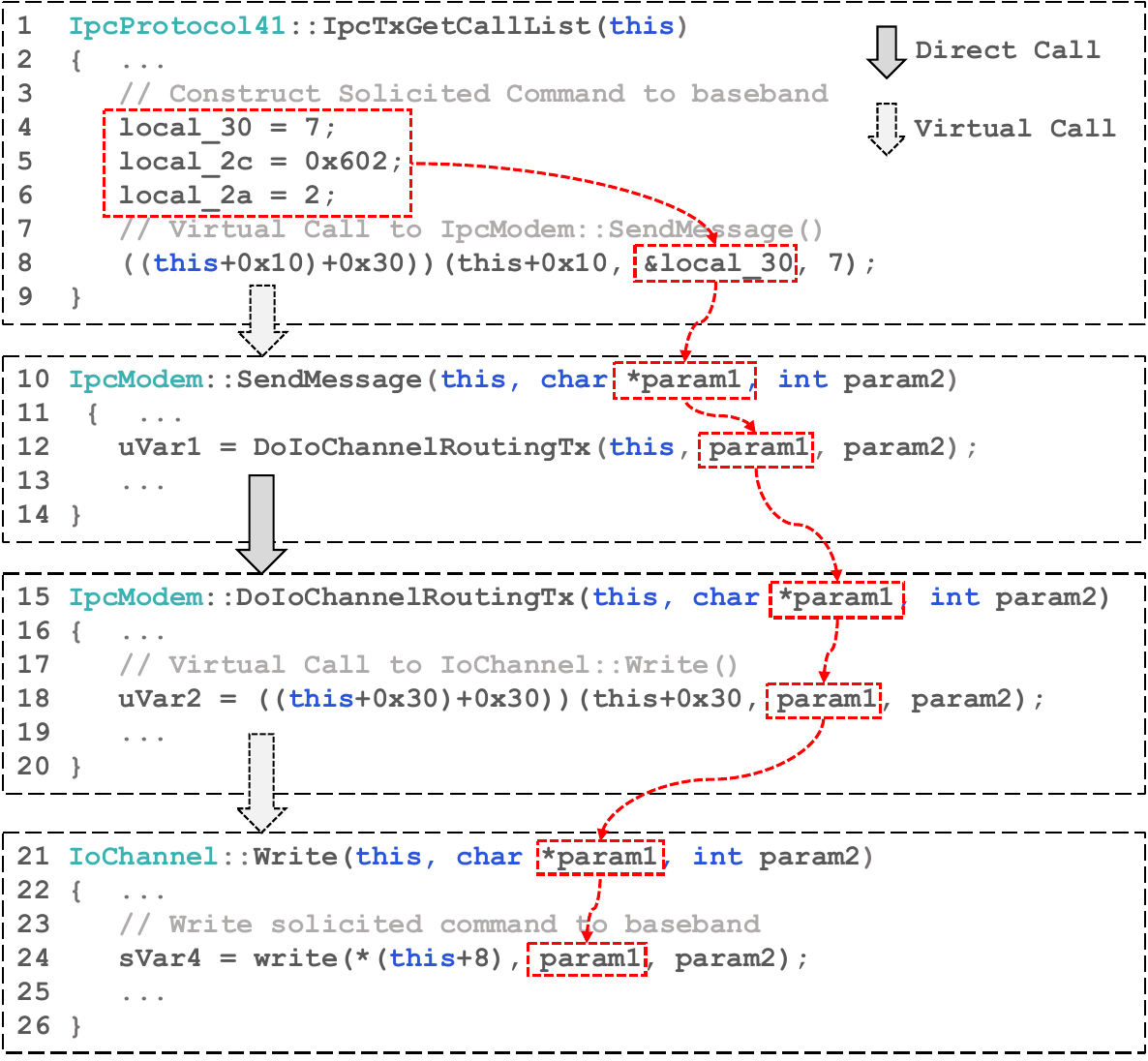}
    \caption{A simplified running example of a solicited command to query call list from the baseband.}
    \label{fig:example}
\end{figure}

\begin{figure*}[t]
    \centering
    \includegraphics[width=0.98\textwidth]{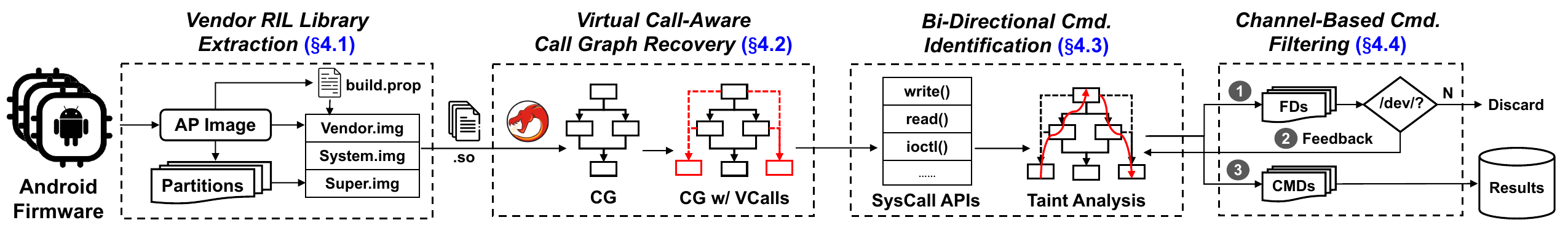}
    \caption{Illustration of the key components of \sys.}
    \label{fig:design}
\end{figure*}

\subsection{Running Example}


To demonstrate our approach, we use a running example: a solicited command from the AP's vendor RIL querying information from the baseband. This example shown in \autoref{fig:example} uses decompiled code from the \texttt{libsec-ril.so} library, extracted from a Samsung Galaxy A53 
with an Exynos chipset. Here, the \texttt{IpcTxCallGetCallList} function constructs the command payload, indicating a request for the call list from the baseband. A virtual function call at line 8 in \texttt{IpcTxCallGetCallList} initiates the command's journey to the baseband. This call uses a function pointer combined with the \texttt{this} pointer and offsets, directing control to the \texttt{SendMessage} function, where the payload is passed as an argument. For simplification, this example does not explain how the virtual callee is resolved and the explanation is given in \S\ref{sec:design2} when we describe how virtual calls are resolved. Following a similar process, the command reaches the \texttt{IoChannel::Write} function through another virtual call and is finally sent to the baseband via the Linux system call API \texttt{write}. 
This example focuses on a solicited command, but the process for unsolicited commands, like AP polling requests from the CP, is similar. Unsolicited commands are typically retrieved from \texttt{read}-like system calls and processed by appropriate utility functions. 

\subsection{Challenges and Insights}

While the example explains how a specific baseband command is constructed and sent, we need to further generalize these observations to achieve our reverse engineering goals. However, we notice there are three major challenges:

\paragraph{(1) Systematically identifying baseband commands}
A mobile device vendor usually defines hundreds 
of specific baseband commands for various radio functionalities, leading to very complicated implementations. For instance, a vendor RIL library we collected typically occupies a few megabytes of binary code with more than 15K functions. Therefore, manual analysis is extremely inefficient, and thus an automated and systematic approach is highly desired. Such a binary analysis approach should precisely locate the baseband commands and correctly extract their syntax. Simply relying on program heuristics (e.g., function and variable symbols) is not scalable and lacks generality, as it will likely introduce errors when encountering different firmware.

To address this challenge, we observe that the baseband commands within the RIL are issued through standard Linux system calls, such as \texttt{write} and \texttt{read} as in the running example in \autoref{fig:example}. Therefore, we design a unified static taint analysis approach~\cite{allen1970control,yang2012leakminer} to precisely track how baseband commands are constructed from these system calls.
Note that we favor a static program analysis approach over dynamic analysis due to scalability concerns, as dynamic analysis requires real devices or emulation capabilities. As a static analysis approach, we only require the RIL binary code, which is readily available from the mobile device firmware.


\paragraph{(2) Efficiently recovering virtual function calls}
The first critical step of static taint analysis is to construct a complete Inter-Procedural Call Graph (CG)~\cite{ming2012ibinhunt}. An incomplete CG will interrupt the correct program control flow and thus prevent our taint analysis from reaching the desired command locations. In particular, virtual function calls (or \textit{vcalls}) are the major roadblocks as current off-the-shelf program analysis frameworks~\cite{ghidra,ida,shoshitaishvili2016sok} cannot resolve vcalls by default during call graph construction. These vcalls are ubiquitous in our problem domain due to the flexibility of programming (e.g., various command handlers invoked from a single statement). For example, in \autoref{fig:example}, the backward command propagation will break at \texttt{IoChannel::Write} if we miss the desired virtual function call (\texttt{IpcModem::DoIoChannelRoutingTx} to \texttt{IoChannel::Write}) from the function call graph.

There have been a handful of program analysis techniques for analyzing virtual function calls at source code level~\cite{bacon1996fast} and recovering virtual inheritance at binary level~\cite{erinfolami2020devil,pawlowski2017marx,elsabagh2017strict}. Our challenge differs in that we are directly solving the concrete vcalls instead of the virtual inheritance. More specifically, when encountering a virtual call site during taint analysis, we need to find the possible callees to propagate the control flow. To this end, we devise a lightweight vcall recovery algorithm that combines class type analysis and offset computation to identify the possible vcall targets and extend the program's control flow.


\paragraph{(3) Accurately filtering undesired commands}
After identifying the possible baseband commands using static taint analysis, we still need to filter these results since there could be many false positives. This is due to the generic static analysis sources and sinks (e.g., the \texttt{write} system call function) that are not only used for baseband commands but also other undesired purposes. For instance, we notice that some RIL implementations use \texttt{read} and \texttt{write} over a local file to achieve asynchronous communications. Therefore, we must accurately filter out undesired commands from our results.

Our key insight comes from the general RIL architecture design in \S\ref{sec:background_ril} that actual interactions between AP and CP are through the Linux-mounted device interfaces, which are distinguished from other types of operations. For example, we notice that both Samsung and Google devices' RIL implementations leverage the mounted devices under the \texttt{/dev/} folder. Those devices represent the communication channels to the peripheral devices, including the baseband. Based on this insight, we leverage the command channel represented by the file descriptor (FD) parameters in the \texttt{read} and \texttt{write} system calls to perform filtering. At a high level, we use backward static analysis to track how the channel FD is initialized and further resolve its path in the system.


%% file: section/4-system.tex
\section{Design and Implementation}
\label{sec:sys}

This section presents the design and implementation of \sys, a static analysis tool to identify and extract the baseband commands from the vendor RIL library. Its high-level workflow is shown in \autoref{fig:design} with four major components. First, taking a vendor-specific Android firmware as input, it automatically unpacks the firmware into multiple partitions and extracts the vendor RIL libraries (\S\ref{sec:design1}). Second, it disassembles the library and produces a virtual call-aware inter-procedural call graph (\S\ref{sec:design2}). Third, based on the enhanced CG, it performs bidirectional taint analysis to uncover both solicited and unsolicited commands (\S\ref{sec:design3}). Lastly, it filters the commands based on their associated communication channels, and those that interact with the baseband are output as results (\S\ref{sec:design4}).

\subsection{Vendor RIL Library Extraction}
\label{sec:design1}

The initial input of \sys is a batch of Android firmware images implemented by various vendors. \sys uses an automated script to unpack the firmware, examine the partitions, and extract the vendor RIL libraries for later analysis. Specifically, it first extracts the AP image and discards the rest (e.g., CP image), and further divides the AP image into multiple partitions. Due to the diversity of the firmware versions we may handle, it targets three standard Android partitions: \texttt{system}, \texttt{vendor}, and \texttt{super}, which contain system-of-chip (SoC) specific code and libraries. Afterward, each of these partitions will be automatically mounted as file systems for library extraction.

The remaining challenge is to accurately extract the desired vendor RIL libraries of interest. Fortunately, while the naming convention and library path are varied across vendors, we find that there is a systematic way to achieve this goal. Specifically, the \texttt{build.prop} file under each partition defines critical system properties and settings, including an important field \texttt{vendor.rild.libpath} that points to the exact vendor RIL library location. As such, our script easily pinpoints and extracts the vendor RIL library.


\subsection{Virtual Call-Aware Call Graph Recovery}
\label{sec:design2}

\begin{figure}[t]
    \centering
    \includegraphics[width=0.40\textwidth]{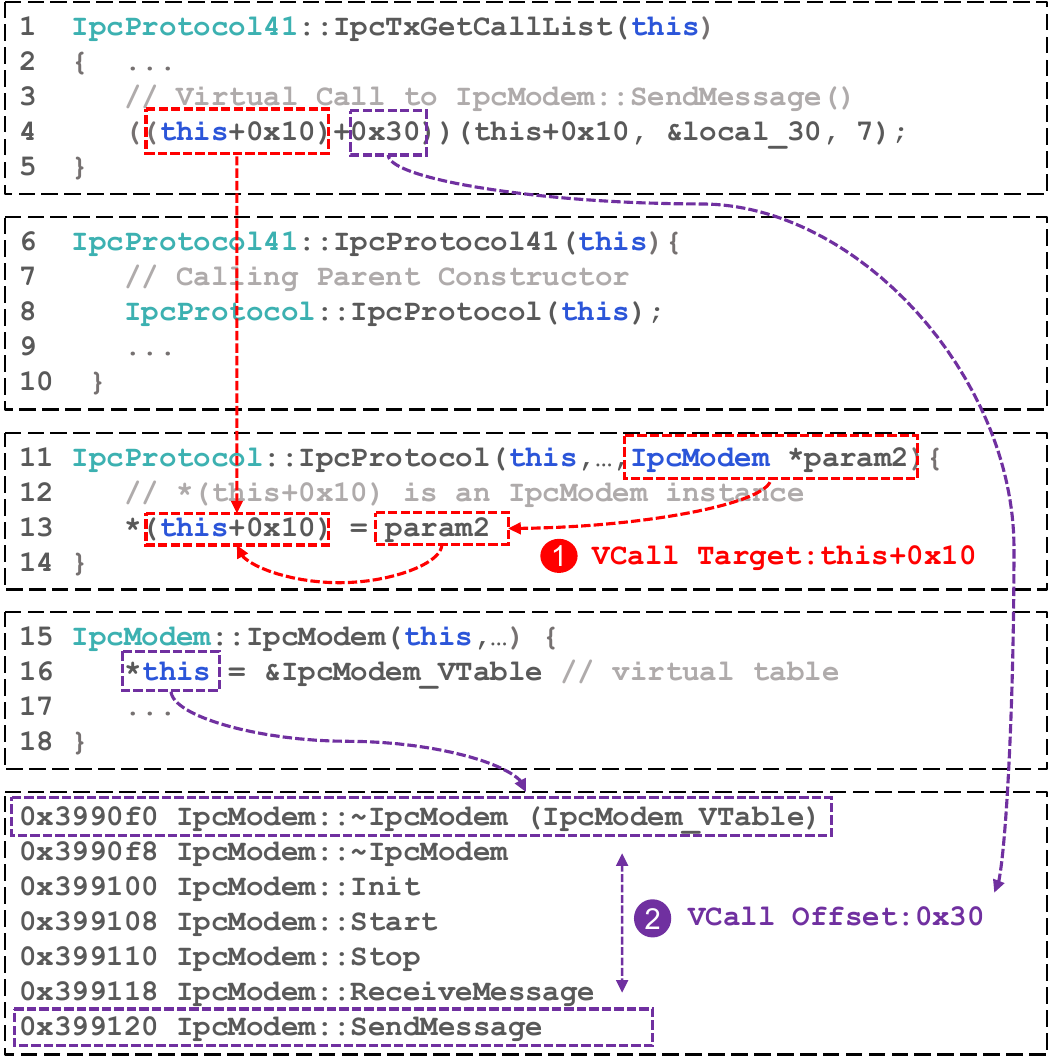}
    \caption{Illustrative example of a virtual function call of our interest in a real vendor RIL library.}
    \label{fig:vcall}
\end{figure}

Given a vendor RIL library, \sys disassembles it and generates an inter-procedural call graph (CG). We use the default disassembler from the state-of-the-art static analysis framework \textsc{Ghidra}~\cite{ghidra}. However, as mentioned in the challenges, the produced CG is incomplete and we still need to recover indirect function calls especially \textit{Virtual Function Calls}. To provide an in-depth explanation, we provide a zoom-in view of the running example (\autoref{fig:example}) with more details about the \textit{Virtual Function Calls}.

\autoref{fig:vcall} explains how \texttt{IpcTxCallGetCallList} from the example can invoke a virtual call to \texttt{IpcModem::SendMessage}. Line 4 reveals the structure of a virtual call which constitutes a virtual call target and a virtual call offset. More specifically, a virtual call target indicates the concrete C++ Class of the instance that initiates the call, and a virtual call offset determines the exact virtual function of that Class. In this example, the virtual call target defined by \texttt{this+0x10} is initialized at the parent constructor function (Line 13 of the \texttt{IpcProtocol} constructor), as \texttt{IpcProtocol41} inherits from \texttt{IpcProtocol}. Based on the statement, the virtual call target is initialized by \texttt{param2} which suggests a pointer to an \texttt{IpcModem} instance based on the parameter type. This further indicates that \texttt{this+0x10} (assigned by \texttt{param2}) also points to an \texttt{IpcModem} instance, and thus the virtual function to be invoked is the \texttt{IpcModem} instance's address pluses a constant offset \texttt{0x30}. To resolve this function, we notice that the \texttt{IpcModem} instance initiates its \texttt{this} pointer to the base of its \textit{Virtual Call Table}. Therefore, we can compute the exact virtual function's address based on the constant offset, and it is thus resolved as \texttt{IpcModem::SendMessage}. Based on the above example, we present a lightweight virtual call solving algorithm in Algorithm~\ref{alg:vcall}. It is broken down into two major tasks to resolve a virtual call as described below.

\input{table/vcall}

\paragraph{(I) Identifying Virtual Call Instructions}
Algorithm~\ref{alg:vcall} takes the control flow graph $\mathbb{G}=(V,E)$ of the vendor RIL library as input, and expands the set of edge $E$ with virtual calls. Since our analysis involves taint analysis that may traverse $\mathbb{G}$ in a backward order, $E$ must be complete to ensure the coverage of analysis (while in forward taint analysis, we can solve only encountered virtual calls). 
Line~\ref{line:vcall_init} to Line~\ref{line:vcall_1} describes the initial steps to identify virtual call instructions of our interest. Fundamentally, virtual calls in disassembly code are of similar structures as indirect calls~\cite{lu2019does,van2016tough,pawlowski2017marx}, as shown in the \autoref{fig:vcall} example. Based on this observation, the algorithm traverses the instructions of each basic block in $V$. It leverages the \textsc{Ghidra}'s P-Code Intermediate Representation (IR) lifter to convert the disassembly into unified P-Code IR syntax as the binaries of interest are cross-architecture~\cite{pcode}. Based on the P-Code instruction syntax, we consider each instruction $i$ represents a potential virtual function call if it is an indirect call instruction (i.e., with \texttt{CALLIND} opcode in P-Code~\cite{pcode}), and if so we add it to the list $I$.
 \looseness=-1

\paragraph{(II) Resolving Virtual Call Function Target}
Line~\ref{line:vcall_2} to Line~\ref{line:vcall_3} illustrates the steps to resolve the virtual calls for each shortlisted instruction $i$. First, it obtains the corresponding P-Code representation of the vcall. In our running example \autoref{fig:vcall}, the vcall at line 4 of the \texttt{IpcTxCallGetCallList} function is represented by \texttt{((this}\texttt{+0x10)+0x30))}, which is described by an abstract syntax tree (AST) structure in P-Code~\cite{pcode}. It then parses this representation to extract the vcall table base $v_t$ (e.g., \texttt{(this+0x10}) and the vcall table offset $v_o$ (e.g., \texttt{0x30}). Since we notice that a Class may reuse its vcalls, we maintain a set $S$ to store and query all vcalls that have already been resolved. To uniquely identify a vcall as a query key, we construct a tuple ($v_c$, $v_t$, $v_o$) using the concrete Class and vcall representation through the following 
two steps:  

\begin{itemize}
    \item \textbf{Virtual Table Analysis}. The first step in our analysis is to identify the concrete Class for a virtual call (vcall) target, determining its virtual table base. This task is challenging in C++ due to difficulties in locating the Class member variable's data definition, which may not be in the same function using the variable. 
    Fortunately, we find that this problem can be solved by limiting the search scope to a list of search functions ($F$) including all (parent) constructors as well as member functions of the current Class $v_c$, which may initialize the target through the shard \texttt{this} pointer.
    For example, in \autoref{fig:vcall}, the list $F$ includes the parent constructor \texttt{IpcProtocol} of the current Class \texttt{IpcProtocol41}, which involves the exact definition of \texttt{this+0x10}. To implement this searching procedure, we use the abstract syntax tree (AST) representation for P-Code variables~\cite{pcode}. It allows us to find the corresponding data definition by matching the AST representations (e.g., \texttt{this+0x10}).
    %
%
     Based on the data definition, a \texttt{class\_inference} procedure (Line~\ref{line:vcall_4}-Line~\ref{line:vcall_5}) is introduced to further infer the concrete Class $c$. Particularly, if the data definition is a function parameter, the corresponding parameter type will be used as the Class type, such as \texttt{this+0x10} in \autoref{fig:vcall} is assigned an \texttt{IpcModem} type of \texttt{param2}.
    %
     Due to the virtual inheritance in C++, such a Class type can indicate all child Classes (e.g., all Classes that inherit from \texttt{IpcModem}). Therefore, it considers the Class hierarchy and returns all possible Classes, which can be inferred from constructors~\cite{fokin2011smartdec}. 

    \item \textbf{Virtual Function Computation}. Based on the inferred Class of the vcall target, the corresponding virtual function table $c_{vt}$ can be looked up from the static memory region, which includes a list of virtual functions within that Class. By adding the base address of the virtual table $c_{vt}$ with the previously resolved offset $v_o$, the concrete vcall function $c_{vf}$ is resolved. Ultimately, the vcall represented by $(i, c_{vf})$ is added to the control flow $E$. 
\end{itemize}


\input{table/api}

\subsection{Bidirectional Command Identification}
\label{sec:design3}

After \sys recovers a virtual call-enhanced call graph, it performs static taint analysis to identify and extract the concrete command values. Since the commands between AP and CP are either solicited or unsolicited, we employ a bidirectional taint analysis to uncover both commands with different sources and sinks. Based on our observation that the communication between AP and CP must be done via Linux system call APIs, we use these APIs as the sources of taint analysis, as shown in \autoref{tab:taint_api}. In the following, we provide details about our bidirectional analysis. 

\paragraph{Backward Taint Analysis}
As shown in our previous running example in \autoref{fig:example}, solicited baseband commands are delivered by system call APIs such as \texttt{write} from user space to the kernel space. These \texttt{Write}-related APIs are thus the boundary of the data flow that we can trace within the vendor RIL library. Other APIs that represent similar write semantics are also included in \autoref{tab:taint_api}, such as \texttt{ioctl} that manipulates device-specific operations. Therefore, the backward taint analysis algorithm starts from these APIs as sources and traces the data flow or the desired API arguments (e.g., the buffer that carries the command payload). \autoref{fig:example} highlights the inter-procedural data flow for such an example starting from the \texttt{write} API. During the backward control flow transfer across functions, it will query the CG to return both the direct and virtual function calls, to ensure the completeness of our results. To define the ending criterion for the backward taint analysis, we consider the command payload to be either a constant value (for \texttt{ioctl} related functions) or a byte array (for \texttt{Read} and \texttt{Write}-related functions) based on our focused APIs in \autoref{tab:taint_api}.
The command shown in \autoref{fig:example} is a byte array local variable, which is stored on the function's stack frame. To handle this case, the algorithm emulates the function's execution to recover the stack frame values, thereby obtaining the command value based on its length.

\begin{figure}[t]
    \centering
    \includegraphics[width=0.43\textwidth]{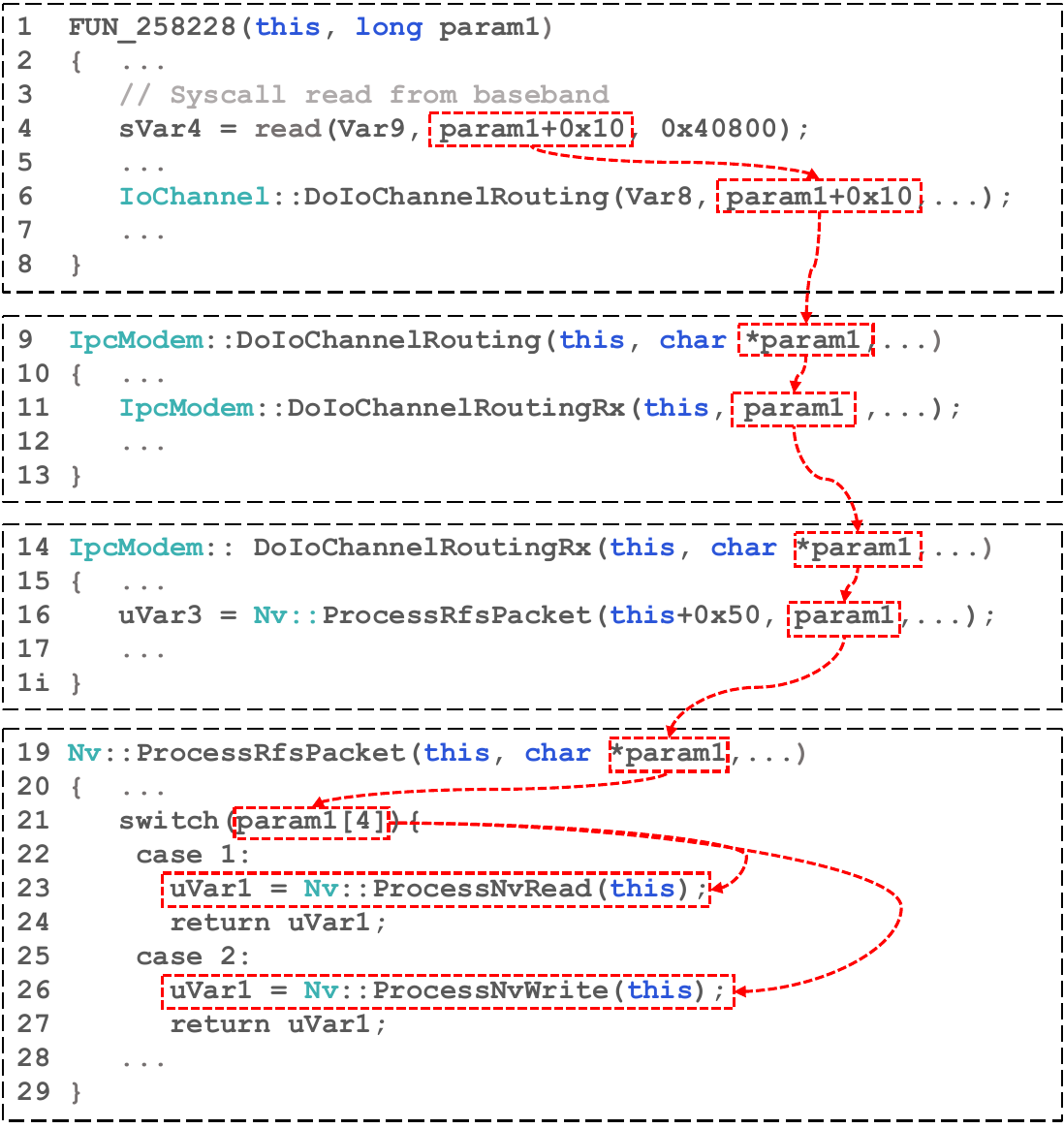}
    \caption{Illustrative example of a forward taint analysis path for an unsolicited baseband command.}
    \label{fig:read}
\end{figure}

\paragraph{Forward Taint Analysis}
In contrast to backward taint analysis, unsolicited commands originating from the CP requires forward taint analysis. To explain this problem, we provide a motivating example in \autoref{fig:read}, in which the AP processes the unsolicited baseband command and invokes different handlers according to the command syntax. Similar to the solicited commands, the unsolicited commands are acquired from the system call API \texttt{read}. They are then stored in a buffer and passed along various processing functions. Eventually, they reach a dispatcher function \texttt{Nv::ProcessRfsPacket} that invokes different handler functions based on the command value. From the descriptions, unsolicited commands are propagated forward after they are read by the vendor RIL. Hence, we define the taint sources to be corresponding \texttt{Read}-related system calls, and the sinks to be comparison instructions (e.g., \texttt{INT\_EQUAL} in P-Code) that represent different branches to dispatch the command.
The forward data flow propagation also considers the virtual function calls recovered in the previous step. 
%

\subsection{Channel-Based Command Filtering}
\label{sec:design4}

\begin{figure}[t]
    \centering
    \includegraphics[width=0.43\textwidth]{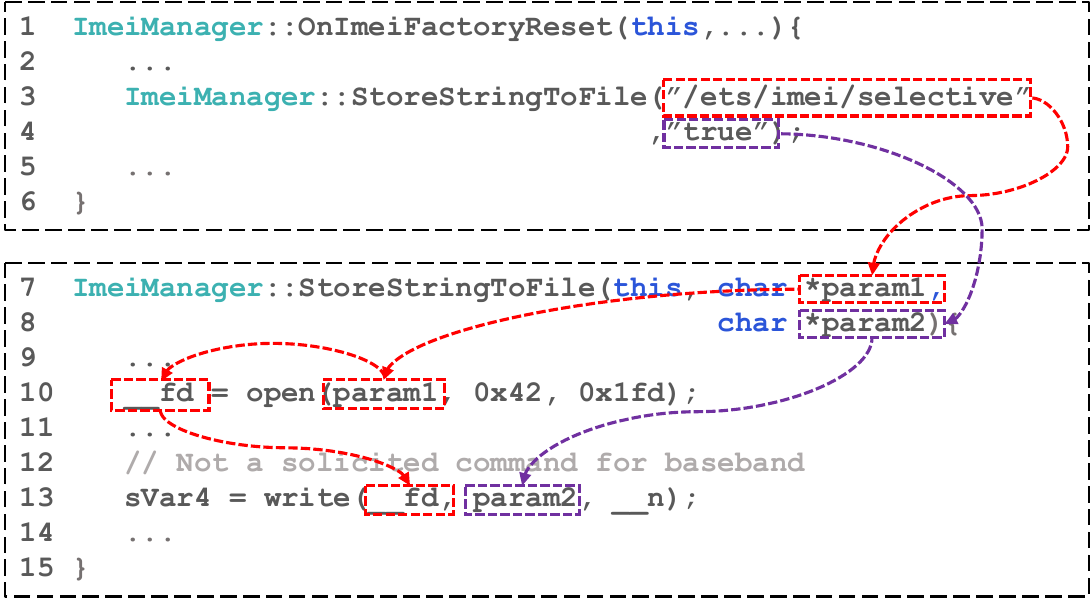}
    \caption{Motivating example of file descriptor-based filtering, showing a \texttt{write} call irrelevant to CP communication.}
    \label{fig:fd}
\end{figure}

Before describing the overall command filtering approach, we explain why it is necessary with a motivating example \autoref{fig:fd}. If \sys simply performs taint analysis on all API call sites, the result represents a superset of the solicited and unsolicited baseband commands, i.e., it includes false positive results which are commands not dedicated to the baseband. The counterexample in \autoref{fig:fd} shows such a case obtained from backward taint analysis from the \texttt{write} system call but turns out to be an undesired command. As shown, the buffer of the \texttt{write} function backward propagates to the function argument of \texttt{StoreStringToFile}, which is invoked to write a string \texttt{true} to a \textit{file descriptor (FD)}. Following the \texttt{\_\_fd} argument, we can pinpoint its definition from the \texttt{open} system call. It is actually an opened file descriptor for a system file at \texttt{/efs/imei/selective}. This example inspires us to use the command channel represented by the file descriptor to filter out false positive results. More specifically, the communication with the baseband is typically conducted through peripheral devices mounted at \texttt{/dev/}, such as USB, audio, and the baseband. Therefore, the system path associated with the file descriptor indicates the communication channel of the system calls. Meanwhile, we assume the vendor RIL library solely handles the baseband communication without engaging with other hardware due to its design, and thus we do not further filter the concrete device names under the \texttt{/dev/} system path.

Next, we describe the detailed algorithm to perform channel-based filtering in Algorithm~\ref{alg:fd}. Before the taint analysis (\S\ref{sec:design3}), it first iterates all targeted system API call sites to filter out those not dedicated to the baseband (Line~\ref{line:fd_init}-Line~\ref{line:fd_1}). To this end, it performs a backward taint analysis to recursively trace the data definition of the file descriptor argument in the targeted system call (e.g., \texttt{read} and \texttt{write}). Since a file descriptor is returned from certain system calls such as \texttt{open} as shown in \autoref{fig:fd}, the taint analysis ends when it reaches a \texttt{CALL} instruction (in P-Code IR~\cite{pcode}) that returns the FD. We further classify this function call into three categories. If it is a \texttt{pipe} system call function, we discard it. If it is a custom function (e.g., a wrapper function of \texttt{open}), it recursively traces the return value in that function. Otherwise, it is an \texttt{open} system call and we switch the analysis target to the file system path argument and perform another backward taint analysis. There could be various system call APIs related to \texttt{open} as we have listed in \autoref{tab:taint_api}. Finally, the path associated with the FD is solved, and the algorithm uses a regular expression to check whether it is a valid path starting with \texttt{/dev/}. It thus filters out those that do not match this criterion and provides feedback to guide the command identification step. Line~\ref{line:fd_2}-Line~\ref{line:fd_3} describes the bidirectional taint analysis procedure that occurs after the system API call site targets an FD that potentially communicates with the baseband.

\input{table/fd}

%% file: table/vcall.tex
\begin{algorithm}[t]
\caption{Virtual Function Call Recovery.}
\label{alg:vcall}
\scriptsize
\SetAlgoLined
\DontPrintSemicolon
\newcommand\mycommfont[1]{\scriptsize\ttfamily\textcolor{black}{#1}}
\SetCommentSty{mycommfont}
\SetKwInOut{Input}{Input}
\SetKwInOut{Output}{Output}
\SetKwComment{Comment}{// }{}
\SetKwProg{Fn}{Function}{}{}
\Fn{\upshape vcall\_recovery($\mathbb{G}$)} { \label{line:vcall_init}
    $(V, E) \leftarrow \mathbb{G}$; $I \leftarrow \emptyset$; $S \leftarrow \emptyset$ \\
    \ForEach{\upshape instruction $i \in V$} {
        \If {\upshape opcode($i$) is indirect call} {
            $I \leftarrow I \cup \{i\}$ \Comment*[r]{Add ins to solved list}
        }
    } \label{line:vcall_1}
    \ForEach{\upshape instruction $i \in I$} { \label{line:vcall_2}
        $v_c \leftarrow $ class of $i$ \;
        $v_t \leftarrow$ virtual call target expression of operand($i$)\;
        $v_o \leftarrow$ virtual call offset of operand($i$)\;
        \If {$(v_c, v_t, v_o) \in S$} {
            continue \Comment*[r]{vcall already solved}
        } 
        $F \leftarrow $ \{$v_c$'s constructors\} $\cup$ \{$v_c$'s class member functions\} \; 
        \ForEach{\upshape function $f \in F$} {
            $c \leftarrow$ class\_inference($v_t$, $f$) \;
            \If {\upshape $c$ is not null} {
                break \;
            }
        }
        $c_{vt} \leftarrow$ $c$'s virtual function table's base\;
        $c_{vf} \leftarrow c_{vt} + v_o$\; 
        $S \leftarrow S \cup \{(v_c, v_t, v_o)\}$ \Comment*[r]{Add to solved list}
        $E \leftarrow E \cup \{(f_i, c_{vf})\}$ \Comment*[r]{Add vcall to CG} \label{line:vcall_3}
    }
}

\Fn{\upshape class\_inference($v_t$, $f$)}{ \label{line:vcall_4}
    $d \leftarrow $ data definition of $v_t$\ in $f$ \;
    \If {\upshape $d$ is a parameter} { 
        \Return parameter's class\;
    }
    \ElseIf {\upshape $d$ is a return variable} {
        \Return return variable type\;
    }
    \ElseIf {\upshape $d$ is a stack variable} {
        \Return class\_inference($d$, $f$)\;
    }
} \label{line:vcall_5}
        
        
\end{algorithm}


%% file: table/api.tex
\begin{table}[t]
    \centering
     \scriptsize
   \scalebox{1.00}{
    \begin{tabular}{l l l r}
        \toprule
        \multirow{2}{*}{\textbf{Analysis Task}} & \multirow{2}{*}{\textbf{API}} & \multirow{2}{*}{\textbf{Parameter Types}} & \multicolumn{1}{c}{\textbf{Tainted}} \\
        & & & \multicolumn{1}{c}{\textbf{Arg Index}} \\ 
        \midrule
        \multirow{4}{*}{Backward Taint Analysis} & {\tt write} & {\tt int}, {\tt void*}, {\tt size\_t} & 0, 1, 2 \\
        & {\tt ioctl} & {\tt int}, {\tt unsigned long} & 1 \\
        & {\tt sendto} & {\tt int}, {\tt void*}, {\tt size\_t}, {\tt int}, ... & 1, 2 \\
        & {\tt \_\_write\_chk} & {\tt int}, {\tt void*}, {\tt size\_t}, {\tt size\_t} & 0, 1, 2 \\
        \midrule
        \multirow{2}{*}{Forward Taint Analysis} & {\tt read} & {\tt int}, {\tt void*}, {\tt size\_t} & 0, 1, 2 \\
        & {\tt \_\_read\_chk} & {\tt int}, {\tt void*}, {\tt size\_t}, {\tt size\_t} & 0, 1, 2 \\
        \midrule
        \multirow{4}{*}{Command Filtering} & {\tt open} & {\tt char*}, {\tt int} & 1 \\
        & {\tt \_\_open\_2} & {\tt char*}, {\tt int} & 1 \\
        & {\tt fopen} & {\tt char*}, {\tt char*} & 0 \\
        & {\tt pipe} & {\tt int} * & 0 \\
        \bottomrule
    \end{tabular}
   }
    \caption{Linux system call APIs for \sys's analysis.}
    \label{tab:taint_api}
\end{table}

%% file: table/fd.tex
\begin{algorithm}[t]
\caption{Channel-Based Command Filtering.}
\label{alg:fd}
\scriptsize
\SetAlgoLined
\DontPrintSemicolon
\newcommand\mycommfont[1]{\scriptsize\ttfamily\textcolor{black}{#1}}
\SetCommentSty{mycommfont}
\SetKwInOut{Input}{Input}
\SetKwInOut{Output}{Output}
\SetKwComment{Comment}{// }{}
\SetKwProg{Fn}{Function}{}{}
\Fn{\upshape command\_filtering($\mathbb{G}$)} { \label{line:fd_init}
    $\mathbb{C} \leftarrow \emptyset$ \Comment*[r]{Final command set}
    \ForEach{\upshape target system API call site $s$} {
        \Comment{Filter commands with fd analysis result}
        $i_{f} \leftarrow$ argument index of fd in $s$\;
        $f \leftarrow$ taint\_analysis($\mathbb{G}$, $s$, $i_f$, 0)  \Comment*[r]{Backward}
        \If{\upshape opcode($f$) is not \texttt{CALL}} {
            continue \;
        }
        \If{\upshape operand($f$) is \textit{open}} {
            $i_{f} \leftarrow$ file system path arg index\;
            $p \leftarrow$ taint\_analysis($\mathbb{G}$, $f$, $i_s$, 0) \;
            \If{\upshape not match($p$, {\scriptsize ``$\wedge$/dev/([$\wedge$/ ]*)+(/[$\wedge$/ ]*)*?\$'')}}  {
                continue \Comment*[r]{Discard command}
            }
        }
        \ElseIf{\upshape operand($f$) is \textit{pipe}} {
            continue \Comment*[r]{Discard command}
        }
        \Else{
            Recursively find fd source \;
        } \label{line:fd_1}
        \Comment{Run taint analysis to extract commands}
        $i_{b} \leftarrow$ argument indices of cmd buffer in $s$ \; \label{line:fd_2}
            $d_{b} \leftarrow$ taint analysis direction for $s$ \; 
        $c \leftarrow$ taint\_analysis($\mathbb{G}$, $s$, $i_b$, $d_b$) \;
        $\mathbb{C} \leftarrow \mathbb{C} \cup \{c\}$ \Comment*[r]{Add command} \label{line:fd_3}
    }
    \Return $\mathbb{C}$ \;
}
\end{algorithm}

%% file: section/5-eval.tex
\input{table/device}

\section{Evaluation}
\label{sec:eval}

We have implemented a prototype of \sys 
with over 4K lines of Java code based on \textsc{Ghidra}~\cite{ghidra}. In support of open science, we have released its implementation on GitHub.
In this section, we present our evaluation results. Specifically, we aim to answer the following research questions: 
\begin{itemize}
    \item \textbf{RQ1:} How many commands were identified for each firmware?
    \item \textbf{RQ2:} How to verify the accuracy of the reverse engineered commands and what is their accuracy?
    \item \textbf{RQ3:} What are the semantics of the reverse-engineered commands and how they are distributed?
    \item \textbf{RQ4:} How did the commands evolve over time? 

\end{itemize}

\subsection{Experiment Setup}

\noindent\textbf{Firmware Collection.} Our evaluation with \sys necessitated a comprehensive firmware analysis. To achieve this, we present an extensive examination of \allril Samsung firmware images. Samsung was selected due to its status as one of the most popular and extensively used Android smartphone brands and its firmware is readily available (e.g., from third-party repositories such as SamMobile~\cite{SamMobil36:online}). However, \sys's analysis methodology is generic and adaptable to other RIL vendors, as discussed in a later section (\S\ref{sec:discuss}).
The collected firmware samples were chosen to cover a wide range of models, regions, and operating system versions. This is to ensure a broad and detailed understanding of the vendor RIL libraries across a spectrum of mobile devices and firmware variations. 
The firmware metadata is detailed in \autoref{tab:devices}.
For brevity, we will refer to the vendor RIL libraries using the associated device model names in the table (e.g., \textbf{G920} for Galaxy S6). 
 \looseness=-1

\paragraph{Experiment Environment}
Our evaluation involves \allril firmware images, which were rigorously tested on a Ubuntu 20.04 laptop equipped with an Intel i7-10710U processor and 16GB of RAM. \sys ran on top of Ghidra 9.2.2 and OpenJDK 11.0.2, providing a stable and efficient environment for our analysis. This setup was carefully chosen to optimize the performance and accuracy of our firmware evaluation process. With respect to the running time, it took \sys 4.2 hours to complete the analysis for the firmware we analyzed. Note that we enabled three threads so that it can process three binaries in parallel, and ensure optimal memory utilization efficiency. On average, it took \sys nine minutes to finish analyzing a single binary.




\subsection{Command Quantity (RQ1)}



\noindent\textbf{Classification.}
Before presenting the statistics, we first discuss the firmware classification methodology. As shown in \autoref{tab:devices}, the \allril firmware falls into two categories based on their RIL communication protocols, including 8 firmware using the legacy \texttt{Ipc41} protocol and 20 firmware using the latest \texttt{Ipc41X} protocol. We were able to identify these firmware attributes based on the program symbols that could not be stripped. For older firmware, where all commands are uniformly implemented under the Class \texttt{IpcProtocol41} without further detailed differentiation of functional modules. In contrast, newer firmware exhibits a more detailed segmentation of vendor RIL layer functionalities, such as \texttt{IpcProtocol41Call} for Phone Call related functions and \texttt{IpcProtocol41Sms} for Short Message Service related functions. Consequently, we have unified these diverse implementations under \texttt{Ipc41} and \texttt{Ipc41X}. This classification approach facilitates later discussion regarding the categorization of different baseband command semantics.



\paragraph{Statistics}
Based on the concrete command values, we eventually obtained \uniquecmd unique baseband commands from the \allril firmware. These statistics indicate that Samsung heavily reuses these protocol-specific commands among the devices to implement communication between AP and CP. 
However, we also notice diversified discrepancies among the firmware we have evaluated, which are discussed later in \S\ref{label:appendix_compare} in the Appendix. 
In \autoref{tab:devices}, we break down the command statistics according to the taint sources defined in \autoref{tab:taint_api}. Note that we have aggregated the results of \texttt{\_\_write\_chk} with \texttt{write} as they are both considered the \texttt{write} system call, which is also applied to \texttt{\_\_read\_chk} and \texttt{read}. The table indicates that \texttt{write} and \texttt{read} significantly contribute to the commands compared to \texttt{Ioctl} and \texttt{Sendto}. Among these two major categories, \sys discovers more \texttt{Write}-related baseband commands than \texttt{Read}-related ones, indicating the developers tend to integrate more diverse functionalities for AP to CP communication than the opposite direction.
  \looseness=-1


\subsection{Command Accuracy (RQ2)}
\label{sec:accuracy}
\sys has extracted hundreds of commands between AP and CP. However, verifying the accuracy of these results is necessary and a matter of concern. 
Despite the challenge of lacking control over the CP both locally and remotely, the correctness of commands sent from CP to AP is relatively easy to confirm due to the partially open-source nature of the AP-side code. 
Additionally, the community has provided numerous tools for debugging the AP-side code and tracing data. 
In contrast, the CP-side operates as a complete black box, with limited knowledge and lacking tools for tracking its runtime behavior. Therefore, this section primarily introduces our testing methodologies for the commands from AP to CP. 
  \looseness=-1

\begin{figure}[t]
    \centering
    \includegraphics[width=0.4 \textwidth]{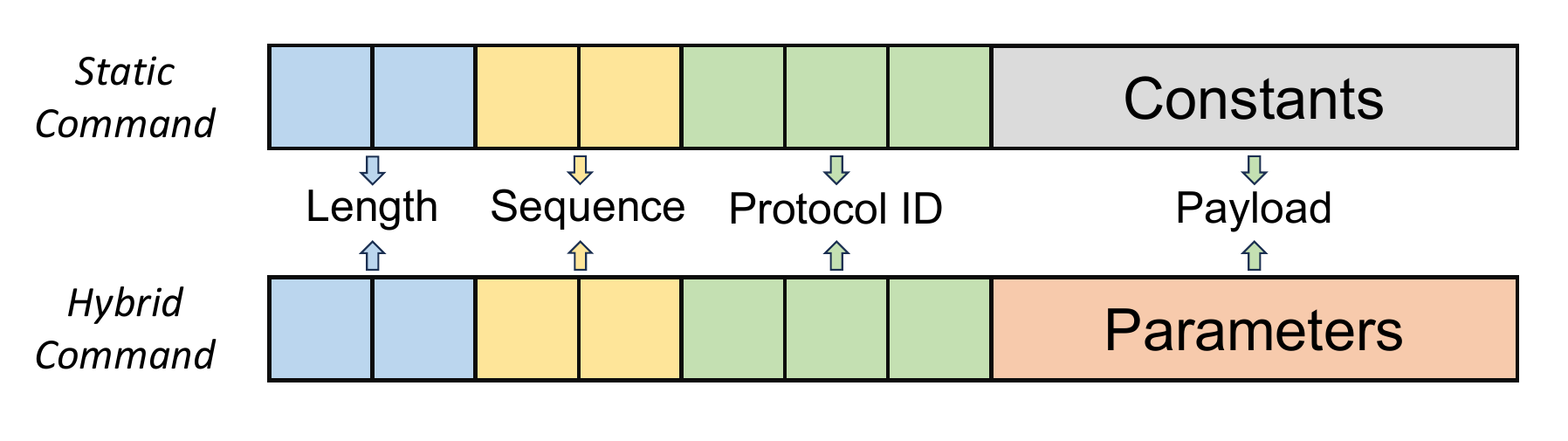}
    \caption{RIL command formats for Samsung Galaxy devices.}
    \label{fig:cmdFormat}
\end{figure}

To validate the commands, we first classify them into two categories: {\bf Static Commands} and {\bf Hybrid Commands}, as shown in \autoref{fig:cmdFormat}. The distinction between the two lies in the fact that hybrid commands include dynamic parameters from the upper layers such as applications, whereas static commands consist of constant values without any parameters. We choose static commands as the dataset for validating the correctness of \sys methodologies because of their parameter-free nature, ensuring the accuracy of the results. However, hybrid commands obtained through static analysis lack runtime parameter information from external sources, and thus cannot be directly validated.

Since these proprietary commands and protocols are not publicly available, it is impossible to compare extracted commands with a ground truth database. To validate the static commands identified, a major challenge is to acquire the command execution results in the modem from the AP side. We observed that after executing the commands, CP sends a response to AP, notifying the execution result.
As such, we use a dynamic analysis approach to trigger real commands and intercept and decode response values to determine the correctness of the uncovered commands. More specifically, we developed a testing framework to automate the sending of static commands through the {\tt /dev/umts\_ipcX} 
interface to CP, triggering a variety of behaviors on a Samsung Galaxy A53 device. Meanwhile, to capture the feedback of command execution, we use \textsc{Frida}~\cite{frida} to hook 
the response handling APIs in RIL (e.g., \texttt{SecRilProxy::OnUnsolicitedResponse}), and check if the response code is in the valid set and whether a non-NULL response has been built.
With this approach, we validated $34\%$ of AP to CP commands on the Samsung Galaxy A53, totaling \allevaluatecmd static commands, with no false positives. For the remaining $66\%$ commands, we further discuss the feasible methodology to validate them in \S\ref{sec:discuss}.

\subsection{Command Semantics (RQ3)}
\label{sec:semantics}

Interestingly, our analysis found that there exist semantic symbols in the RIL binaries, though they are {\it not} required by \sys to perform the analysis. These symbols actually come from the \textit{.dynsym} section and thus cannot be stripped due to dynamic linking purposes. 
To make use of these symbols for command comprehension, we extract the root node function of the taint analysis chain and extract the function name semantics to understand the corresponding command. The intuition is that the root node function that generates the command typically provides the corresponding semantic associated with the command introduced by the developers. As illustrated in \autoref{fig:example}, the root node function in this taint analysis chain, namely \texttt{IpcTxCallGetCallList}, suggests its functionality is to transmit a request to retrieve the list of phone calls. However, other functions along the taint analysis chain, such as \texttt{DoIoChannelRoutingTx}, do not exhibit a direct relationship with the command's functionality.

\input{table/semantics}

Based on the root node function names, we developed an automatic tool for analyzing the command semantics. First, we employ a well-known tokenization technique in Natural Language Processing (NLP), which breaks the function names into meaningful semantic tokens. Next, we perform a classification step to aggregate the command semantics in order to obtain a high-level view. To this end, building upon the generated tokens, we filter out the prefix tokens and verbs, such as \texttt{Ipc} that do not indicate useful command-wise semantics. 
The first non-filterable word encountered, typically a noun, is then assigned as the functional category. For example, the tokens of the function \texttt{IpcTxCallGetCallList} shown in \autoref{fig:example}, are first filtered to exclude \texttt{Ipc}, \texttt{Tx}, and \texttt{Get}. 
Ultimately, the function is classified as belonging to the \texttt{Call} category. We call each distinctive category a \textit{module} in the below discussion.

\paragraph{Semantic Module Distribution}
Utilizing the aforementioned methodology, we categorized distinct commands into corresponding functional modules. Then, we conducted a comprehensive analysis of the command frequencies within these modules, with select results presented in \autoref{tab:semantics}, which delineates the top-10 modules for \texttt{Write} and \texttt{Read}-related commands in \texttt{Ipc41} and \texttt{Ipc41X}, and provide the respective command counts and percentages within each module.
Upon comparing the numerical counts and proportions, it becomes evident that, despite the version discrepancies between \texttt{Ipc41} and \texttt{Ipc41X}, the fundamental functionalities remain largely consistent, aligning with our expectations. 

\paragraph{Semantics-guided Command Analysis}
By semantically clustering commands, we can contrast different command values within the same category, thereby gaining insights into specific bytes of command values. For instance, through comparing command values under the \texttt{Sim} and \texttt{Domestic} modules, we observed that the fifth byte, \texttt{0x05} and \texttt{0x20} respectively, serves as the identifier for their modules. By cross-validating with other modules, we found a consistent pattern where the fifth byte across all commands is utilized to denote the command group. This analysis allows us to reverse engineer the proprietary RIL command format in \autoref{fig:cmdFormat}.
Such information further provides meaningful guidance for dynamic testing such as attack payload discovery discussed later in \S\ref{sec:attack_discovery}. 
%
Additionally, the segmentation of command functionalities facilitates a more streamlined analysis, allowing us to focus on modules characterized by a higher command volume and complex functionalities. It optimizes the identification of vulnerable points and potential security issues. In \S\ref{sec:attack}, we will expound upon the security issues and potential attacks discovered in Samsung firmware based on our comprehension of command semantics.

\subsection{Command Evolution (RQ4)}
\label{sec:diff}

To answer our RQ4, we provide two high-level overviews of the command evolution with respect to time and device series. 



\begin{figure}[t]
    \centering
    \includegraphics[width=0.48\textwidth]{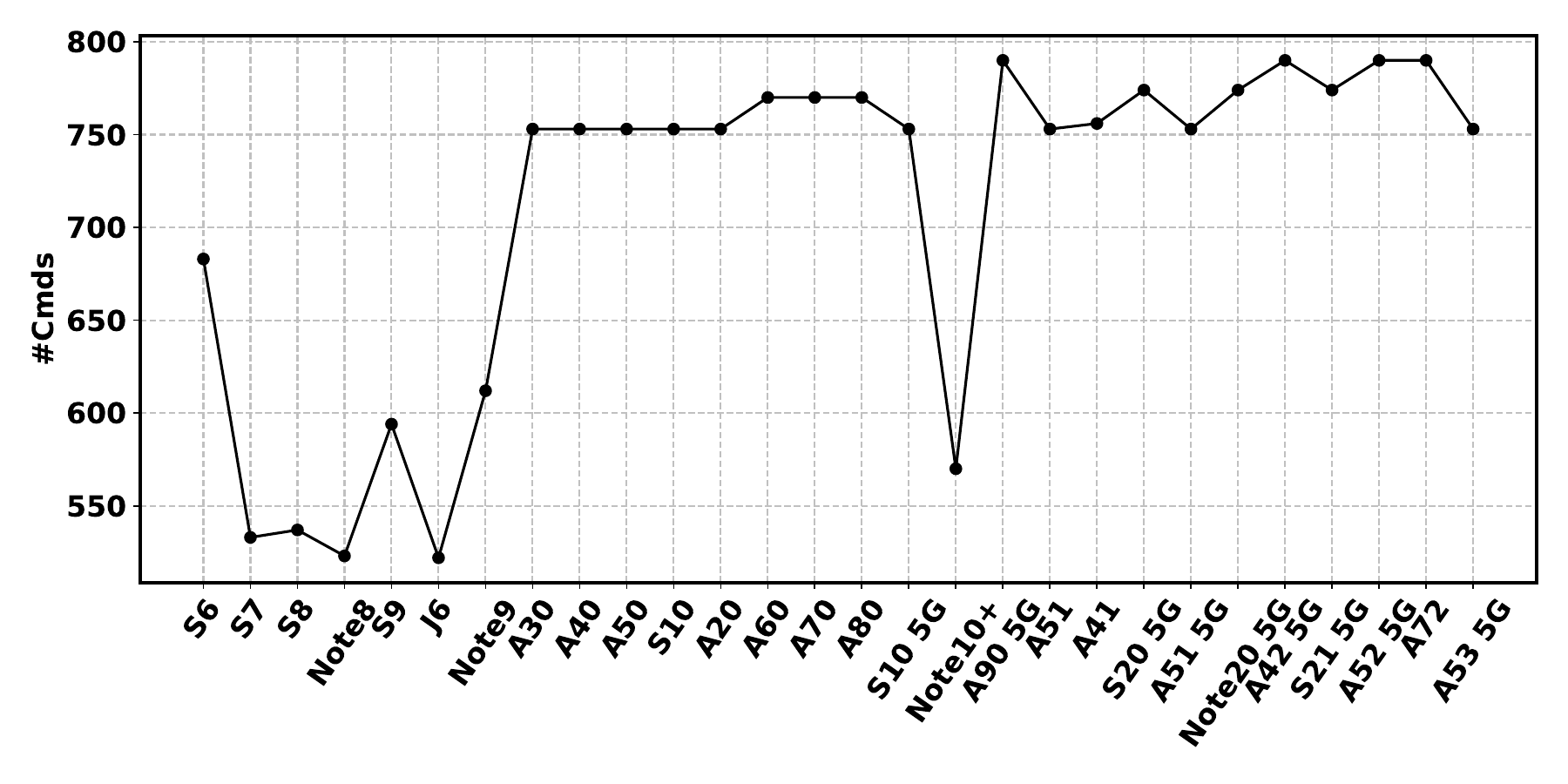}
    \caption{Historical view of baseband command evolution.}
    \label{fig:history}
\end{figure}

\paragraph{Historical View}
We first visualize the evolution of baseband command quantities in \autoref{fig:history}, where the devices are ordered chronologically per their release date (ranging from 2015 to 2022). The figure shows a growing trend in the number of commands over time, indicating that Samsung has continuously integrated new features into the baseband. Notably, we notice a substantial increase (over 20\%) of command numbers after Note 9 was released since the devices afterward have evolved to the \texttt{Ipc41X} protocol that brings in significant new commands. More specifically, we found that the new \texttt{Ipc41X} protocol splits the original functions in \texttt{Ipc41} into more subdivided functional areas, such as \texttt{IpcProtocol41Call} and \texttt{IpcProtocol41Sms}. This also implies the architecture of CP firmware is likely to have been refactored. 
Additionally, 5G device models tend to include more baseband functions than non-5G devices, and there is also one interesting exception (Note 10+) in which \sys only reported less than 600 commands, far less than other devices released in its time frame due to the legacy \texttt{Ipc41} implementation in the 4G version.

\paragraph{Device Series View}
We further provide a different perspective regarding the device series in \autoref{fig:boxplot}. The figure categorizes the results into three device series, namely series A (15 models), series Note (4 models), and series S (8 models). The J series model is excluded since there is only one such firmware in our dataset. Our visualization suggests obvious discrepancies in the number of commands among these three series. Specifically, series A models tend to involve more commands since all of them have evolved to the latest \texttt{Ipc41X} protocol. In contrast, the Note series and S series involve many legacy models that use the old \texttt{Ipc41} protocol.


%% file: table/device.tex
\begin{table*}[t]
  \centering
  \scalebox{0.98}{
  \scriptsize
  \begin{tabular}{llllcllrrrrr}
    \toprule
    \multicolumn{7}{c}{\textbf{Firmware Metadata}} & \multicolumn{5}{c}{\textbf{Evaluation Statistics and Commands per API}} \\
    \cmidrule(lr){1-7} \cmidrule(lr){8-12}
    \textbf{Device Name} & \textbf{Device Model} & \textbf{Firmware} & \textbf{Region} & \multicolumn{1}{l}{\textbf{OS Ver.}} & \textbf{Ipc41} & \textbf{Ipc41X} & \multicolumn{1}{l}{\textbf{\#VCall}} & \multicolumn{1}{l}{\textbf{\#Write}} & \multicolumn{1}{l}{\textbf{\#Read}} & \multicolumn{1}{l}{\textbf{\#Ioctl}} & \multicolumn{1}{l}{\textbf{\#Sendto}} \\
    \midrule
    Samsung Galaxy Note10+	&	SM-N975F	&	N975FXXU1ASHE	&	Egypt	&	9	&	\fullcirc	&	\emptycirc	&	2,828	&	478	&	49	&	41	&	2	\\
    Samsung Galaxy J6	&	SM-J600FN	&	J600FNXXU5BSH1	&	UK	&	9	&	\fullcirc	&	\emptycirc	&	2,476	&	427	&	53	&	40	&	2	\\
    Samsung Galaxy S6	&	SM-G920F	&	G920FXXS6ETI6\_G920FEVR6ETK1\_EVR	&	UK	&	7	&	\fullcirc	&	\emptycirc	&	3,057	&	570	&	67	&	45	&	1	\\
    Samsung Galaxy S7	&	SM-G930F	&	G930FXXU8EUE1\_G930FOXA8EUE1\_BTU	&	UK \& IRE	&	8	&	\fullcirc	&	\emptycirc	&	2,736	&	433	&	53	&	45	&	2	\\
    Samsung Galaxy S9	&	SM-G960F	&	G960FXXUHFVB4\_G960FOXMHFVC2\_TNZ	&	New Zealand	&	10	&	\fullcirc	&	\emptycirc	&	3,020	&	493	&	55	&	44	&	2	\\
    Samsung Galaxy S8	&	SM-G950F	&	G950FXXU5DSFB	&	Romania	&	9	&	\fullcirc	&	\emptycirc	&	2,552	&	445	&	53	&	37	&	2	\\
    Samsung Galaxy Note8	&	SM-N950F	&	N950FXXUFDUG5\_N950FUWEFDUG4\_IUS	&	Mexico	&	9	&	\fullcirc	&	\emptycirc	&	2,551	&	429	&	53	&	39	&	2	\\
    Samsung Galaxy Note9	&	SM-N9600	&	N9600ZHU9FVB3\_N9600OWC9FVB3\_IUS	&	Mexico	&	10	&	\fullcirc	&	\emptycirc	&	3,517	&	493	&	55	&	62	&	2	\\
    Samsung Galaxy A20	&	SM-A205F	&	A205FXXSACVC2\_A205FOLMACUI2\_MM1	&	SINGAPORE	&	11	&	\emptycirc	&	\fullcirc	&	3,300	&	505	&	204	&	42	&	2	\\
    Samsung Galaxy A30	&	SM-A305F	&	A305FDDU6CUI3\_A305FOLM6CUJ1\_MM1	&	SINGAPORE	&	11	&	\emptycirc	&	\fullcirc	&	3,301	&	505	&	204	&	42	&	2	\\
    Samsung Galaxy A40	&	SM-A405FN	&	A405FNXXU4CVD1\_A405FNOXM4CVD1\_EUR	&	SPAIN	&	11	&	\emptycirc	&	\fullcirc	&	3,302	&	505	&	204	&	42	&	2	\\
    Samsung Galaxy A41	&	SM-A415F	&	A415FXXS1CVC1\_A415FOXM1CVA3\_MEO	&	Hungary	&	11	&	\emptycirc	&	\fullcirc	&	3,296	&	505	&	204	&	45	&	2	\\
    Samsung Galaxy A50	&	SM-A505F	&	A505FDDU9CVC2\_A505FOJM9CVC2\_LYS	&	UAE	&	11	&	\emptycirc	&	\fullcirc	&	3,304	&	505	&	204	&	42	&	2	\\
    Samsung Galaxy A60	&	SM-A606Y	&	A606YXXU2CVA1\_A606YOLO2CVA1\_BRI	&	Taiwan	&	11	&	\emptycirc	&	\fullcirc	&	4,549	&	505	&	204	&	59	&	2	\\
    Samsung Galaxy A70	&	SM-A705F	&	A705FXXU5DVC1\_A705FOLM5DVB2\_XXV	&	VIETNAM	&	11	&	\emptycirc	&	\fullcirc	&	4,550	&	505	&	204	&	59	&	2	\\
    Samsung Galaxy A80	&	SM-A805F	&	A805FXXS6DVD1\_A805FOWC6DUJ1\_UNE	&	Mexico	&	11	&	\emptycirc	&	\fullcirc	&	4,549	&	505	&	204	&	59	&	2	\\
    Samsung Galaxy S20 5G	&	SM-G981B	&	G981BXXUDFVC7\_G981BOXMDFVC7\_XSA	&	Croatia	&	12	&	\emptycirc	&	\fullcirc	&	3,327	&	525	&	205	&	42	&	2	\\
    Samsung Galaxy S21 5G	&	SM-G991B	&	G991BXXS4CVCG\_G991BOWO4CVB9\_IUS	&	Mexico	&	12	&	\emptycirc	&	\fullcirc	&	3,327	&	525	&	205	&	42	&	2	\\
    Samsung Galaxy Note20 5G	&	SM-N981B	&	N981BXXS3FVC8\_N981BOXM3FVC5\_NZC	&	UK \& IRE	&	12	&	\emptycirc	&	\fullcirc	&	3,327	&	525	&	205	&	42	&	2	\\
    Samsung Galaxy A42 5G	&	SM-A426B	&	A426BXXU3DVC2\_A426BOXM3DVC2\_MM1	&	ITV	&	12	&	\emptycirc	&	\fullcirc	&	4,580	&	525	&	204	&	59	&	2	\\
    Samsung Galaxy A52 5G	&	SM-A526B	&	A526BXXS1CVD1\_A526BOLM1CVB6\_XSA	&	Australia	&	12	&	\emptycirc	&	\fullcirc	&	4,580	&	525	&	204	&	59	&	2	\\
    Samsung Galaxy A72	&	SM-A725F	&	A725FXXU4BVC1\_A725FOJM4BVC1\_ILO	&	ISRAEL	&	12	&	\emptycirc	&	\fullcirc	&	4,580	&	525	&	204	&	59	&	2	\\
    Samsung Galaxy A90 5G	&	SM-A908B	&	A908BXXU5EVD2\_A908BOXM5EVD2\_AUT	&	Switzerland	&	12	&	\emptycirc	&	\fullcirc	&	4,578	&	525	&	204	&	59	&	2	\\
    Samsung Galaxy A51	&	SM-A515F	&	A515FXXU5FVD2\_A515FOLM5FVD2\_MM1	&	SINGAPORE	&	12	&	\emptycirc	&	\fullcirc	&	3,323	&	526	&	189	&	36	&	2	\\
    Samsung Galaxy A51 5G	&	SM-A516B	&	A516BXXU5DVC2\_A516BOXM5DVC2\_AUT	&	SPAIN	&	12	&	\emptycirc	&	\fullcirc	&	3,323	&	526	&	189	&	36	&	2	\\
    Samsung Galaxy S10	&	SM-G973F	&	G973FXXUEHVC6\_G973FOWCEHVC6\_IUS	&	Mexico	&	12	&	\emptycirc	&	\fullcirc	&	3,320	&	526	&	189	&	36	&	2	\\
    Samsung Galaxy S10 5G	&	SM-G977B	&	G977BXXSBHVD1\_G977BOXMBHVC6\_BTU	&	UK \& IRE	&	12	&	\emptycirc	&	\fullcirc	&	3,325	&	526	&	189	&	36	&	2	\\
    Samsung Galaxy A53 5G	&	SM-A536E	&	A536EXXS4AVJ3\_A536EOWO4AVI2\_ARO	&	Argentina	&	12	&	\emptycirc	&	\fullcirc	&	3,327	&	526	&	189	&	36	&	2	\\
  \bottomrule
  \end{tabular}
  }
  \caption{Metadata of the Android firmware and their evaluation statistics ($\fullcirc$ indicates the firmware belongs to that category).}
  \label{tab:devices}
\end{table*}

%% file: table/semantics.tex
\begin{table*}[t]
    \centering
     \scriptsize
    \begin{tabular}{clrrp{3cm}|clrrp{5cm}}
        \toprule
        \multicolumn{5}{c|}{\textbf{Write-related Commands}} & \multicolumn{5}{c}{\textbf{Read-related Commands}} \\ 
        \midrule
        \textbf{Protocol} & \textbf{Module} & \textbf{\#} & \textbf{\%} & \textbf{Semantics} & \textbf{Protocol} & \textbf{Module} & \textbf{\#} & \textbf{\%} & \textbf{Semantics} \\ 
        \midrule
         \multirow{10}{*}{Ipc41} & Domestic & 332 & 8.81\% & Domestic & \multirow{10}{*}{Ipc41} & Nv & 152 & 34.70\% & Non-Volatile \\
         & Cfg & 256 & 6.79\% & Configuration &  & Net & 17 & 3.88\% & Network \\
         & Net & 240 & 6.37\% & Network &  & Sms & 17 & 3.88\% & Short Message Service \\
         & Call & 213 & 5.65\% & Call &  & Qmi & 16 & 3.65\% & Qualcomm MSM Interface \\
         & Ss & 146 & 3.87\% & Supplementary Services &  & Ss & 15 & 3.42\% & Supplementary Services \\
         & Snd & 110 & 2.92\% & Sound &  & Call & 9 & 2.05\% & Call \\
         & Ims & 102 & 2.71\% & IP Multimedia Subsystem &  & Cdma & 9 & 2.05\% & Code Division Multiple Access \\
         & Factory & 91 & 2.42\% & Factory &  & Disp & 9 & 2.05\% & Display \\
         & Qmi & 82 & 2.18\% & Qualcomm MSM Interface &  & Embms & 9 & 2.05\% & Evolved Multimedia Broadcast Multicast Service \\
         & LTE & 68 & 1.80\% & Long-Term Evolution &  & Factory & 9 & 2.05\% & Factory \\
         \midrule
         \multirow{10}{*}{Ipc41X} & Domestic & 860 & 8.31\% & Domestic & \multirow{10}{*}{Ipc41X} & Cfg & 660 & 16.47\% & Configuration \\
         & Net & 740 & 7.15\% & Network &  & Domestic & 580 & 14.47\% & Domestic \\
         & Cfg & 700 & 6.77\% & Configuration &  & Nv & 380 & 9.48\% & Non-Volatile\\
         & Call & 420 & 4.06\% & Call &  & Call & 265 & 6.61\% & Call \\
         & Ss & 300 & 2.90\% & Supplementary Services &  & Sec & 160 & 3.99\% & Security \\
         & Snd & 260 & 2.51\% & Sound &  & Net & 140 & 3.49\% & Network \\
         & Factory & 260 & 2.51\% & Factory &  & Sap & 120 & 2.99\% & SIM Access Profile \\
         & Qmi & 220 & 2.13\% & Qualcomm MSM Interface &  & Sat & 120 & 2.99\% & Serving and Tracking Area \\
         & Ims & 212 & 2.05\% & IP Multimedia Subsystem &  & Ss & 120 & 2.99\% & Supplementary Services \\
         & Sim & 180 & 1.74\% & Subscriber Identity Module &  & Sim & 100 & 2.50\% & Subscriber Identity Module \\
        \bottomrule
    \end{tabular}
    \caption{Baseband command semantics categorized by top-10 modules for different protocols.} 
    \label{tab:semantics}
\end{table*}

%% file: section/6-attack.tex
\section{Attacks via RIL}
\label{sec:attack}
To evaluate the security implication of the extracted commands, this section presents how we are able to construct attack payloads and demonstrate attacks based on our results and findings in \S\ref{sec:eval}. We present an automatic attack payload discovery framework, followed by examples of attacks initiated from either the AP or CP on a Samsung A53 5G device (model SM-A536E in \autoref{tab:devices}).

\begin{figure}[t]
    \centering
    \includegraphics[width=0.26\textwidth]{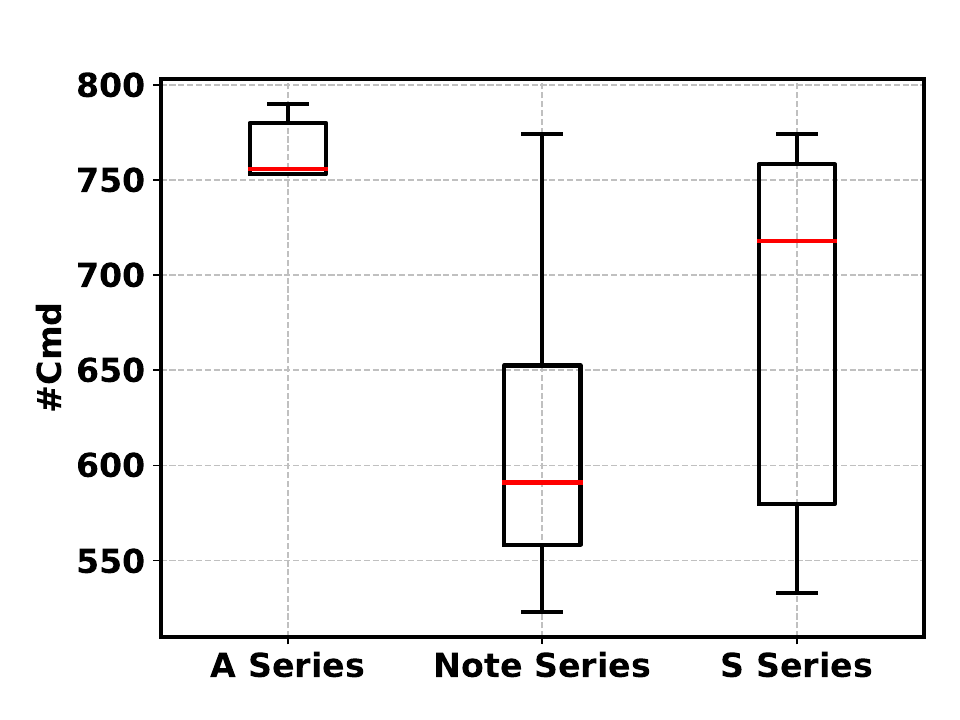}
    \caption{Device series view of command statistics.}
    \label{fig:boxplot}
\end{figure}

\begin{figure}[t]
  \centering
  \begin{subfigure}[t]{0.23\textwidth}
    \centering
    \includegraphics[height=94pt]{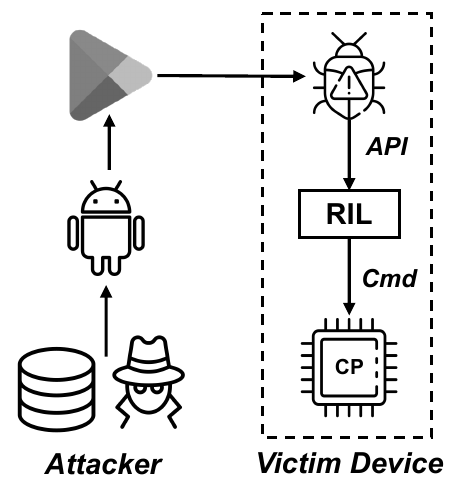}
    \caption{Attacks targeting CP.}
    \label{fig:writeattack}
  \end{subfigure}
  \hfill
  \begin{subfigure}[t]{0.23\textwidth}
    \centering
    \includegraphics[height=94pt]{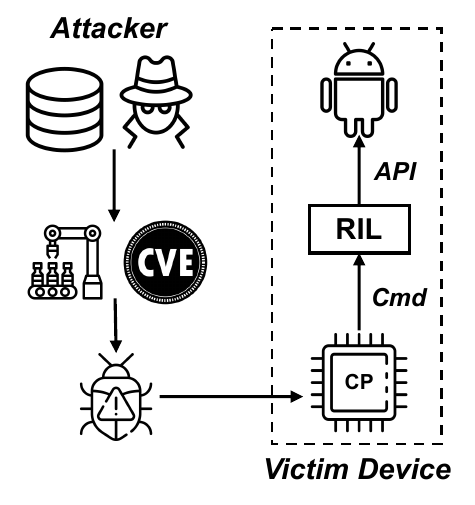}
    \caption{Attacks targeting AP.}
    \label{fig:readattack}
  \end{subfigure}
  \caption{Workflow of two attacks exploiting the baseband commands from our reverse engineered results.}
  \label{fig:attack}
\end{figure}

\paragraph{Threat Model}
We consider that the identified commands from the vendor RIL libraries could be exploited in various ways to compromise user security and privacy. First, regarding {\it AP to CP} attackers (shown in \autoref{fig:writeattack}), they could inject malicious solicited baseband commands through RIL to cause substantial impact such as service disruption without the user's consent. To this end, we consider three distinctive adversary classes with different privileges:


\begin{enumerate}[leftmargin=12pt]
    \item \textbf{Privileged Attackers}. Since the command injection through RIL is protected by Android's permission mechanism by default~\cite{android_ril}, we first consider privileged attackers who have root access or are within the radio group. Such an attacker model may be achieved by examples such as a Remote Access Trojan (RAT)~\cite{android_rat} or malicious system applications (e.g., a dialer app).
    
    
    \item \textbf{Unprivileged Attackers Exploiting Unprotected APIs}. Interestingly, we found that the interaction with RIL can be achieved by unprivileged attackers. This adversary class involves any unprivileged user-space applications, which can be readily introduced to the victim device through app stores and social engineering. These applications further escalate their privileges by exploiting unprotected Android framework APIs, such as secret menus~\cite{android_hidden_code}, backdoors, and app-accessible system function calls (e.g., \texttt{invokeOemRilRequestRaw})~\cite{shao2016kratos}, which allows the injection of arbitrary RIL payloads to baseband.
    
    \item \textbf{Unprivileged Attackers Exploiting Unprotected Communication Interfaces}. An alternative way for a user-space malicious program to escalate its privilege is through unprotected communication interfaces. One notable example is the unprotected UNIX sockets implemented in Samsung Galaxy devices that allow any applications to interact with the RILD and inject arbitrary RIL messages to the CP~\cite{samsung_rild}.
    
\end{enumerate}

Regarding {\it CP to AP} exploitation (depicted in \autoref{fig:readattack}), it requires an adversary with high privilege, e.g., a compromised baseband, kernel, or the OS, that injects malicious RIL commands to the AP through RIL. This could be achieved through existing RCE exploits~\cite{zeng2012design,golde2016breaking,hernandez_firmwire_2022,maier2020basesafe} or supply chain adversaries. 
%
%


\subsection{Attack Payload Discovery}
\label{sec:attack_discovery}
As discussed in \S\ref{sec:accuracy}, observing the behavior of commands sent from AP to CP is much more challenging than the opposite direction.
Consequently, based on the identified command in \S\ref{sec:eval}, we develop an automatic black-box testing framework to discover and validate command payloads exploitable for AP to CP attacks, based on existing debugging tools such as the Android Debug Bridge (ADB). We then applied this framework to a Samsung Galaxy A53 device. In particular, we focus on crash-related attacks, as they can be easily observed with practical security implications.

Our framework extends the testing tool in \S\ref{sec:accuracy} with two additional modules: \textbf{States Checking Module} and \textbf{Parameters Mutation Module}. Specifically, in the States Checking Module, we compare the mobile device's state before and after command execution, checking if ADB is connectable and if the baseband modem is still in service, to assess the potential malicious exploitation of certain commands. Additionally, inspired by popular fuzzing works~\cite{fioraldi2020afl++, bohme2017directed, li2022muafl, shi2024invivo, li2021para}, the Parameters Mutation Module mutates bytes representing run-time parameters within hybrid commands and uses various response values and status as guidance, to identify values triggering anomalies on the device.

\paragraph{States Checking Module}
The States Checking Module quires the \texttt{telephony.registry} interface's return value, specifically \texttt{mServiceState}. This process determines whether the modem is affected by the command, and assesses the extent of impact.
After injecting a command to CP, it checks the state of the modem immediately. If the modem recovers shortly after a transient \texttt{OUT\_OF\_SERVICE} state, it categorizes the command as a \textit{Temporary Crash Command}, indicating that it could at least disrupt the modem's service shortly. 
Subsequently, the device is rebooted, and the modem state is queried again. If the modem has returned to an \texttt{IN\_SERVICE} state, the module labels this command as a \textit{Recoverable Crash Command}, indicating that manual intervention can restore service. However, if the modem remains in an \texttt{OUT\_OF\_SERVICE} state, indicating that the command has permanently damaged the firmware's functionality, it is classified as a \textit{Permanent Crash Command}.

\paragraph{Parameters Mutation Module}
For hybrid commands, we port AFL++~'s seed mutation mechanism~\cite{fioraldi2020afl++} to mutate their corresponding parameter bytes. Additionally, we utilize response values and modem states as guidance to indicate interesting mutation results.
\input{table/vulcmds}




\subsection{Attacks Targeting CP}
\label{sec:attackcp}

\noindent\textbf{Attack Goal and Procedure.} 
As shown in \autoref{fig:writeattack}, the attacker aims at compromising the CP through a malicious AP application via RIL, such as disrupting the cellular service and performing denial-of-service attacks. To begin with, the attacker needs to acquire sufficient knowledge about the attack command payload for the victim device. The desired attack payload could either be static RIL commands that trigger expected outcomes (e.g., power off the baseband), or hybrid commands with malicious parameters (e.g., to set invalid baseband channels). For instance, by running the automated attack payload discovery tool, we show that an attacker can identify 7 exploitable RIL commands (in \autoref{tab:vulCmds}) for a Samsung Galaxy A53 model, which takes two weeks to complete. To launch the attack, the attacker implants malicious logic as a privileged or user-space application to be installed on the victim device, which injects the malicious commands to the baseband at run-time.

\paragraph{Security Implication}
\autoref{tab:vulCmds} summarizes the 7 commands that can be leveraged to compromise a Samsung Galaxy A53 baseband, which leads to different degrees of impact based on our repeated experimentation. Among them, 3 commands can cause \textit{Temporary Crashes} to the baseband, which recovers after a few seconds. In order to persist the attacks, one can constantly replay these attack payloads to cause service disruption. In addition, we found one command that causes \textit{Recoverable Crash}, and the victim will have to reboot the device to obtain cellular service. In extreme cases, we found three commands that trigger \textit{Permanent Crashes}, which requires a complete firmware re-flash to recover the baseband. We provide the exploit details in \S\ref{sec:appendix_details_cp} of Appendix.



\paragraph{Root Cause Analysis}
We speculate the root causes for the aforementioned attacks are two-fold. First, static RIL commands are essentially abused for malicious purposes as they represent inherent baseband functionalities. For instance, in \autoref{tab:vulCmds}, we list the corresponding command semantics extracted from the symbols, including two static commands that reset and power off the baseband. Moreover, for hybrid RIL commands, we speculate that the CP-side parameter checks for command parameters are insufficient, leading to crashes (e.g., segmentation faults) on the CP. Based on the command structures and semantics, these hybrid commands tend to encapsulate complex parameters (e.g., APDU and SMS payloads) that require sophisticated decoding logic to handle them properly.



\subsection{Attacks Targeting AP}
\label{sec:attackap}

\noindent\textbf{Attack Goal and Procedure.} 
Based on the analysis of \texttt{Read}-related commands, we have identified that the \texttt{Nv} Module contains a set of functions accessing the file system on the AP, such as \texttt{ProcessNvWrite} and \texttt{ProcessNVRead}. In the presence of a compromised baseband modem, these baseband functionalities can be exploited for unauthorized file system access from the high-privileged CP. To this end, the baseband firmware code may initiate a malicious unsolicited command request to the AP via RIL, by embedding the target file path as the command parameter. Notably, while this vulnerability was first reported in 2014 as a backdoor~\cite{samsung_backdoor}, it still exists in a widespread of the device firmware we have analyzed. 
Meanwhile, we discovered an undisclosed vulnerability in \texttt{Nv::ProcessOpenFile} due to the absence of symbolic link checking, which further allows arbitrary file access on the AP.

\paragraph{Security Implication}
Typically, due to the invocation of vendor RIL libraries by the \texttt{RILD} process, which possesses high system privileges, read and write operations are granted to the attacker on the majority of files. 
In an existing exploit demonstration by the Replicant project~\cite{samsung_backdoor}, it is shown that an attacker can inject malicious logic to the modem kernel driver (equivalent to an attacker CP), which is capable of opening and reading a user file located under the AP's \texttt{/data/} folder. To exploit this vulnerability further, our new finding shows that a malicious symbolic link could be provided as input to enable arbitrary file access on the AP. We provide additional details about this vulnerability in \S\ref{sec:appendix_details_ap} of Appendix.

\paragraph{Root Cause Analysis}
The unsolicited commands within the \texttt{Nv} module are possibly designed for the CP to read and write certain logs to the AP's file system, such as dumping the baseband logs. However, since the command allows to specify the file paths within the AP, it enables arbitrary file system access from a malicious CP. Moreover, we also explored the possibility of using \texttt{../} for directory traversal attacks (to escape the prefix paths) and found that the firmware has eliminated them with corresponding sanitizer logic. \looseness=-1




%% file: table/vulcmds.tex


\begin{table*}[t]
\centering
 \footnotesize
\begin{tabular}{llll}
\toprule
\textbf{Crash Type}        & \textbf{Command Semantics}      & \textbf{Cmd Type} & \textbf{Command Payload (Bytes)} \\
\midrule
\multirow{3}{*}{Temporary}                  & IpcProtocol41Power::IpcTxResetOemModem                & Static & {\tt 07, 00, 00, 00, 01, 03, 05}  \\
                           & IpcProtocol41Sap::IpcTxGetSapTransferApdu           & Hybrid & {\tt 0e, 01, 00, 00, 12, 04, 02, ...}  \\
                           & IpcProtocol41Imei::IpcTxImeiPreconfigSet             & Hybrid & {\tt 17, 00, 00, 00, 10, 03, 03, ...}  \\
\midrule
Recoverable                & IpcProtocol41Power::IpcTxModemPowerOff                & Static & {\tt 07, 00, 00, 00, 01, 02, 01}  \\
\midrule
\multirow{3}{*}{Permanent}                  & IpcProtocol41Net::IpcTxSetSystemSelectionChannels   & Hybrid & {\tt 20, 00, 00, 00, 08, 07, 03, 01, ff (x8) ...}  \\ 
                           & IpcProtocol41Domestic::IpcTxDomesticSetChannelSettingLte & Hybrid & {\tt 09, 00, 00, 00, 20, 64, 03, ...}  \\
                           & IpcProtocol41Domestic::IpcTxDomesticGetNsriDecryptSms    & Hybrid & {\tt 99, 00, 00, 00, 20, 91, 02, ...} \\
\bottomrule
\end{tabular}
    \caption{RIL command payloads uncovered by the attack discovery framework and the crash types they could trigger.}
\label{tab:vulCmds}
\end{table*}

%% file: section/7-discussion.tex
\section{Discussion}
\label{sec:discuss}

\paragraph{Coverage of Validated Commands}
As discussed in \S\ref{sec:accuracy}, a substantial portion ($66\%$) of the baseband commands discovered involve parameters from external sources, which cannot be directly validated via our framework. Examples of such include heterogeneous inputs from user-space applications, file systems, and networks, such as SMS, configuration files, as well as APDU payloads. These hybrid commands would be useful for further security analysis. For instance, the structural information of these parameter values is useful for 
dynamic analysis, such as the attack payload discovery process discussed in \S\ref{sec:attack_discovery}, thereby enabling us to identify more exploitable commands. Further, a grey-box fuzzing tool could be developed with parameter-aware mutation mechanisms.  However, it is challenging to construct these command payloads with valid parameters. 
In the following, we categorize them into three types and discuss feasible ways to validate them in future work and show the corresponding examples in ~\autoref{sec:appendix_details_hybrid}. 
\begin{enumerate}[leftmargin=12pt]
    \item \textbf{Direct Input Parameter}. \autoref{fig:hybrid1} shows an example of a parameter directly from an external input to configure the baseband. These parameters are typically of primitive types (e.g., boolean and integer) and are easy to handle, as they directly contribute to the command payload through simple operations (e.g., concatenation). We have extended \sys and our fuzzer to support this type of hybrid command by marking payload bytes as static or dynamic during taint analysis and validated an additional $16\%$ of AP to CP commands on the Samsung Galaxy A53.  \looseness=-1
    
    \item \textbf{Derived Parameter}. As shown in \autoref{fig:hybrid2}, a derived parameter needs to be transformed (e.g., through splitting, reassembly, or conditional evaluation) before it is used as part of the command payload. As these parameters involve computation and are condition-dependent, we consider using symbolic execution~\cite{shoshitaishvili2016sok} to track each byte of the parameter and determine its constraints. Once the constraints are identified, fuzzing can be employed to explore the correct commands. 
    
    \item \textbf{Structured Parameter}. We also found that many parameters are constructed as certain data structures. An example in \autoref{fig:hybrid3} shows a JSON payload involving network scanning information to be transmitted to the baseband. Due to the complexity of these data structures and lack of constraint conditions in JSON files, direct exploration using fuzzing would be highly challenging. Therefore, a more feasible way is to first obtain input seeds for the fuzzer, such as dynamically intercepting valid command payloads from normal traffic via Frida. However, such an approach needs to properly trigger these interactions, which still remains a challenge.
\end{enumerate}
%




%

\paragraph{Generality of Methodology}
While in this paper we focus on Samsung's RIL libraries, \sys's methodology is generic and could be applied to other Android RIL vendors such as Qualcomm and MediaTek. This is due to the standard RIL architecture in Android as presented in \autoref{fig:ril_arch}~\cite{android_ril}, where the vendor RIL libraries must rely on universal Linux system calls (e.g., \texttt{read} and \texttt{write}) to interact with the baseband regardless of the vendors. We consider extending \sys for other device vendors as an important future work to unveil additional security issues across a diverse range of devices on the market.


\paragraph{Responsible Disclosure}
We have reported our findings to Samsung Security, providing the exploit prerequisites and Proof of Concept (PoC). 
As of this writing, Samsung has confirmed our findings and patched the vulnerabilities, and awarded us a bug bounty for the lack of symbolic link checking bug within their RIL implementation.
%

%% file: section/8-related.tex
\section{Related Work}
\label{sec:related}

\noindent\textbf{Baseband Security.}
The baseband, managing vital cellular functions yet being closed-source and complex, holds significant security interest for both attackers and researchers. There have been several remote code execution (RCE) exploits due to baseband memory corruption \cite{grassi2018exploitation,grassi2021over,weinmann2012baseband}, accompanied by guides on analyzing and debugging baseband code \cite{golde2016breaking,berard2020design}. Recently, automated static analysis has been used to detect deviations in baseband code from cellular standards \cite{kim2023basecomp,kim2021basespec,hernandez2019basebads}, uncovering memory bugs. Furthermore, advancements in dynamic firmware analysis and rehosting have facilitated baseband fuzzing \cite{maier2020basesafe,hernandez_firmwire_2022,hussain2021noncompliance}. However, research on the RIL remains limited. Previous studies mainly leveraged RIL for cellular traffic monitoring \cite{vallina2013rilanalyzer,li2016mobileinsight} and attack mitigation \cite{wen2023thwarting}. \textsc{ARIstoteles} represents a closely related work that scrutinizes Apple's baseband interfaces and uses fuzzing to identify security flaws~\cite{kroll2021aristoteles}. However, our research targets Android RIL, a completely different ecosystem that allows baseband vendor-specific implementations. 


\ignore{
\noindent\textbf{Baseband Security.}
Baseband is of high-security value to both attackers and security researchers as it handles critical cellular functions but is closed-source and extremely complex. There have been multiple remote code execution (RCE) exploits through baseband memory corruption~\cite{grassi2018exploitation,grassi2021over,weinmann2012baseband} as well as tutorials to analyze and debug the baseband code~\cite{golde2016breaking,berard2020design}. More recently, automated static analysis techniques are proposed to identify the deviation of baseband code from cellular specifications~\cite{kim2023basecomp,kim2021basespec,hernandez2019basebads}, which have also revealed memory corruption bugs. Recent efforts have also advanced dynamic firmware analysis and rehosting techniques to support baseband fuzzing~\cite{maier2020basesafe,hernandez_firmwire_2022,hussain2021noncompliance}. In contrast, the Radio Interface Layer (RIL) is a relatively unexplored area. Prior research mainly uses RIL for cellular traffic monitoring~\cite{vallina2013rilanalyzer,li2016mobileinsight} and attack prevention~\cite{wen2023thwarting}. \textsc{ARIstoteles} uses static analysis to vet Apple's baseband interfaces and performs fuzzing to uncover insecure implementations~\cite{kroll2021aristoteles}. In this paper, we propose a novel approach to reverse engineer baseband from the RIL.  
}

\paragraph{Mobile Security}
%
Orthogonal to our work, other prior research has delved into various critical components of the mobile system firmware. A few studies target the pre-installed applications in Android OS firmware, uncovering privilege escalation~\cite{elsabagh2020firmscope} and over-the-air update vulnerabilities~\cite{blazquez2021trouble}. Some extract proprietary AT commands and use fuzzing to understand their security implications~\cite{tian2018attention,karim2019opening}. Note that the commands we reverse engineered are not AT commands as they do not have the prefix \texttt{AT}~\cite{karim2019opening}. In addition, vendor-specific private APIs~\cite{el2021dissecting} and access policies~\cite{hernandez2020bigmac} have been identified and raised concerns regarding access control vulnerabilities. To comprehensively measure the Android firmware security, some large-scale longitudinal studies were conducted~\cite{hou2022large,possemato2021trust}.

\ignore{
AT command analysis~\cite{tian2018attention} and fuzzing~\cite{karim2019opening} 
~\cite{elsabagh2020firmscope} pre-installed apps, privilege escalation in Android firmware
~\cite{blazquez2021trouble} Firmware over-the-air apps, static analysis, privacy, key management
~\cite{el2021dissecting} OEM private APIs in custom Android ROM
~\cite{hou2022large,possemato2021trust} security measurement Android ROM firmware
~\cite{hernandez2020bigmac} Android firmware policy analysis
}

\paragraph{C++ Binary Analysis}
Binary analysis of C++ has been challenging due to various sophisticated mechanisms. Previous work mainly tackles class hierarchy reconstruction~\cite{pawlowski2017marx,erinfolami2019declassifier}, class and object recovery~\cite{schwartz2018using,jin2014recovering}, and employing static analysis to uncover bugs and vulnerabilities~\cite{schubert2019phasar,wen2023egg}. In particular, virtual inheritance is a unique challenge in object-oriented languages such as C++. There are techniques to recover virtual inheritance~\cite{erinfolami2020devil} and defenses to protect the control flow integrity of virtual function calls~\cite{pawlowski2019vps,elsabagh2017strict}. \sys addresses the challenge of recovering virtual function calls and employs static taint analysis for command extraction.


%% file: section/9-conclusion.tex
\section{Conclusion}
\label{sec:conclude}

This paper presents a novel approach to unveil security vulnerabilities of basebands from the Radio Interface Layer in the Android ecosystem. Based on this insight, we develop \sys, an automatic tool employing static binary code analysis to reverse engineer vendor-proprietary baseband commands from the vendor RIL binaries. It addresses multiple technical challenges of binary analysis including the recovery of C++ virtual function calls as well as the comprehensive identification and accurate filtering of baseband commands. We have applied \sys to \allril vendor RIL libraries for Samsung mobile devices, and identified \uniquecmd unique baseband commands. To evaluate the security implications of our results, we further demonstrated \totalattack exploitable attacks with the discovered commands on a Samsung Galaxy A53 device, which can interfere with the baseband's cellular service and allow arbitrary access to the AP's file system.

%% file: section/10-appendix.tex
\appendix






\section{Comparative Analysis}
\label{label:appendix_compare}
Based on the commands identified by \sys, we executed a comparative analysis to pinpoint command discrepancies across devices, with results detailed in \autoref{tab:diff}. For this, we selected firmware from two specific models as baselines. As shown in \autoref{tab:diff}, the sections above and below the horizontal line represent class \texttt{Ipc41} and \texttt{Ipc41X} respectively, with N975 and A536 firmware serving as baselines. Our baseline selection aimed to ensure balanced representation in the comparative data. We introduce two terms for this analysis: \textbf{Base Unique} (\textbf{BaseUni}), indicating commands present in the baseline but missing in the current firmware; and \textbf{Current Unique} (\textbf{CurUni}), denoting commands in the current firmware but absent in the baseline. Notably, the semantics of extracted commands can also be understood from the semantic symbols from the \textit{.dynsym} section described in \S\ref{sec:semantics}, and thus we also make use of these related semantics in the remaining discussion.

\begin{packeditemize}
\item \textbf{Discrepancies in \texttt{Ipc41} Firmware}.
Our analysis reveals significant variation among the \texttt{Ipc41} firmware. In terms of \texttt{Write}-related commands, each firmware had 12 to 179 unique commands compared to the baseline. Many of these commands relate to \texttt{Call} and \texttt{Domestic} configurations, differing across devices. Notably, the G920 firmware displayed the most variance, with 179 unique commands, due to an additional \texttt{IpcProtocol40} not present in other versions. This difference is linked to the Galaxy S6, the oldest device in our study, using Android 7. For \texttt{Read}-related commands, we observed less variation, with 5 to 22 unique commands per firmware. 

   \item \textbf{Discrepancies among \texttt{Ipc41} firmware}.   The comparative analysis results exhibit a notable degree of heterogeneity among the \texttt{Ipc41} firmware. Regarding \texttt{Write}-related commands, each firmware presents 12 to 179 exclusive commands from the baseline. Among these unique commands, we found they are associated with semantic keywords indicating \texttt{Call} and \texttt{Domestic} configurations which vary across devices. It is noteworthy that the G920 firmware stands out by incorporating a substantial divergence, featuring as many as 179 unique commands not found in other variants. This discrepancy arises due to the implementation of an additional protocol, \texttt{IpcProtocol40}, within the G920 firmware, a feature absent in other firmware. We link the exceptional phenomenon to the corresponding device Galaxy S6 in \autoref{tab:devices}, noteworthy for being the sole and oldest device using Android 7. Regarding \texttt{Read}-related commands, we found each firmware differs in possessing 5 to 22 commands unique to the baseline firmware, representing less heterogeneity compared to \texttt{Write}.

\item \textbf{Discrepancies in \texttt{Ipc41X} Firmware}. 
In contrast to \texttt{Ipc41}, the \texttt{Ipc41X} firmware showed reduced disparity in \texttt{Write} commands and more concentrated differences. Here, the primary unique command was \texttt{SetNrModeConfig}, used for 5G New Radio (NR) configurations in newer models. This indicates a stabilization in Samsung's \texttt{Write}-related baseband functions in recent devices. However, for \texttt{Read} commands, there was a noticeable increase in unique commands, with up to 20 new ones, mostly related to \texttt{Call} and \texttt{Display} functions.

\end{packeditemize}

\input{table/diff}

\section{Vulnerability and Exploit Details}
\label{sec:appendix_details}
Based on the result of \sys, we conducted an analysis of the commands between AP and CP. Consequently, we identified potential security risks for both AP and CP. In the following sections, we will elaborate on these vulnerabilities and potential attacks.

\subsection{Attacks Targeting CP}
\label{sec:appendix_details_cp}
\begin{figure}[h]
    \centering
    \includegraphics[width=0.32\textwidth]{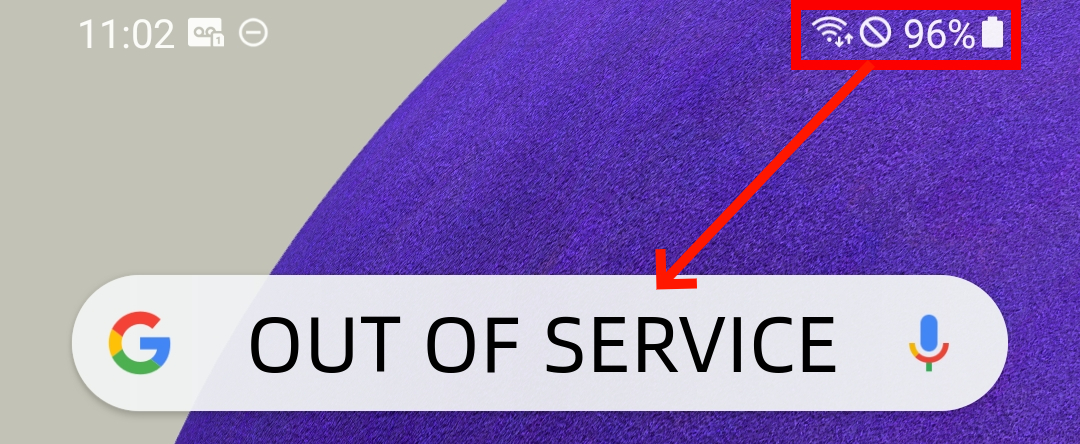}
    \caption{A screenshot indicating a crashed baseband of a Samsung Galaxy A53 after attack command injection.}
    \label{fig:modemcrash}
\end{figure}

As depicted in \autoref{code:autoAttack}, we automate the procedure of command testing and payload replay on the Android side via the \texttt{ADB} interface from the host machine. Furthermore, as described in \S\ref{sec:attack_discovery}, we query the modem status at different points in time to provide feedback on command execution. The command injection tool, \texttt{test\_cmd}, as illustrated in \autoref{code:testCmd}, is utilized for simulating command launches from AP to CP. With this tool, we format the payloads related to write commands obtained from automated reverse engineering of RIL and then transmitted them to the \texttt{/dev/umts\_ipcX} interface. If using the attack payload, shown in \autoref{tab:vulCmds}, they will result in various levels of modem crash, notably the \texttt{OUT\_OF\_SERVICE} state from the state query response. This can also be observed visually from the Android OS's status bar as depicted in \autoref{fig:modemcrash}.

\definecolor{mGreen}{rgb}{0,0.6,0}
\definecolor{mGray}{rgb}{0.5,0.5,0.5}
\definecolor{mPurple}{rgb}{0.58,0,0.82}
\definecolor{backgroundColour}{rgb}{0.95,0.95,0.92}
\lstset{
language=C,
basicstyle=\scriptsize\ttfamily,
numbers=left,
numberstyle=\tiny,
frame=lrtb,
columns=fullflexible,
showstringspaces=false,
backgroundcolor=\color{backgroundColour},   
commentstyle=\color{mGreen},
keywordstyle=\color{magenta},
numberstyle=\tiny\color{mGray},
stringstyle=\color{mPurple},
breakatwhitespace=false,         
breaklines=true,                 
captionpos=b,                    
keepspaces=true,                 
numbers=left,                    
numbersep=5pt,                  
showspaces=false,                
showstringspaces=false,
showtabs=false,                  
tabsize=2,
xleftmargin=5pt,
}

\begin{table}[t]
\begin{center}
\begin{minipage}{\linewidth}
\begin{lstlisting}[language=Python, frame=single, numbers=left, linewidth=8cm, caption={Automatic attack script.}, label={code:autoAttack}]
# auto_attack.py
test_cmd = ["adb shell su -c /data/local/tmp/test_cmd"]
dump_cmd = ["dumpsys telephony.registry"]

def dump_sys():
    stdout = su_exec_adb_command(dump_cmd)
    lines = stdout.splitlines()
    for line in lines:
        if "mServiceState" in line:
            return contain_bad_state(line)

payload_files = get_files_absolute_paths("./payload")
crashed = False
for payload in payload_files:
    dump_alert = dump_sys()
    if crashed:
        if dump_alert:
            print("Permanent Crash at previous cmd")
            exit(-1)
        else:
            print("Recoverable Crash at previous cmd")
            crashed = False
    push_command = ['push', payload, '/data/local/tmp/']
    exec_adb_command(push_command)
    change_payload_name(payload, "attack_payload.txt")
    test_process = subprocess.Popen(test_cmd)
    test_process.wait()
    dump_crash = dump_sys()
    if dump_crash:
        crashed = True
        print("Temporary Modem Crash")
    root_cmd = ['reboot']
    exec_adb_command(root_cmd)
\end{lstlisting}
\end{minipage}
\end{center}
\end{table}

\begin{table}[t]
\begin{center}
\begin{minipage}{\linewidth}
\label{code:testCmd}
\begin{lstlisting}[language=C,linewidth=8cm, caption={Attack payload testing script.}, label={code:testCmd}]
// test_cmd.c
#define MAX_BYTE_DATA_SIZE 2048
int main() {
    FILE *hex_file = fopen("attack_payload.txt", "r");
    char hex_string[MAX_BYTE_DATA_SIZE];
    fgets(hex_string, sizeof(hex_string), hex_file);
    fclose(hex_file);
    unsigned char byte_data[MAX_BYTE_DATA_SIZE / 2];
    char *token;
    int counter = 0;
    token = strtok(hex_string, ", ");
    while (token != NULL) {
        sscanf(token, "%hhx", &byte_data[counter++]);
        token = strtok(NULL, ", ");
    }
    int fd;
    char filename[] = "/dev/umts_ipc0";
    fd = open(filename, O_WRONLY, 0666);
    write(fd, byte_data, sizeof(byte_data));
    close(fd);
    printf("Attack send to modem successfully.\n");
    return 0;
}
\end{lstlisting}
\end{minipage}
\end{center}
\end{table}

\subsection{Attacks Targeting AP}
\label{sec:appendix_details_ap}

In addition to identifying potential security risks of commands from AP to CP through automation, we manually discovered vulnerabilities in CP's access to the AP-side file system through RIL by understanding the semantics of the reverse-engineered commands.

\begin{figure}[t]
    \centering
    \includegraphics[width=0.37\textwidth]{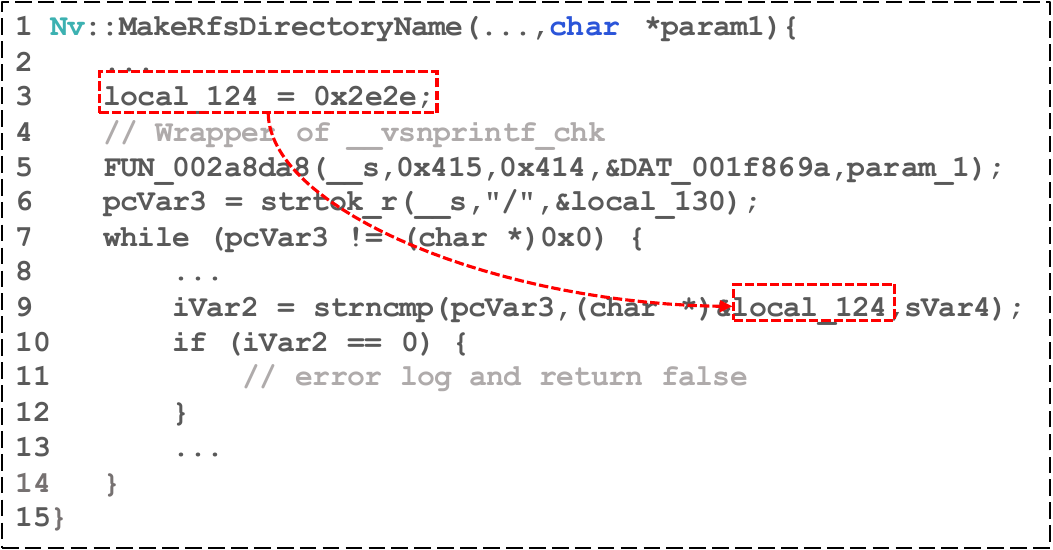}
    \caption{Soft link check absence in NV Open File.}
    \label{fig:2echeck}
\end{figure}

The Vendor RIL offers the Nv class for CP to manipulate the AP's file system via commands. Within the \texttt{Nv::MakeRfsDirectoryName} function called by \texttt{Nv::ProcessOpenFile}, there exists a check for the string ".." as shown in \autoref{fig:2echeck}, which could be exploited in directory traversal attacks. However, there is no verification for whether the incoming path is a symbolic link. An attacker, controlling both the CP and cooperative AP-side symbolic links, could execute arbitrary file read and write operations.

\section{Hybrid Baseband Commands}
\label{sec:appendix_details_hybrid}

In addition to the descriptions in \S\ref{sec:discuss}, we provide more details about the three types of hybrid baseband commands with examples to illustrate how they are constructed and used in the RIL.

\begin{figure}[t]
    \centering
    \includegraphics[width=0.4\textwidth]{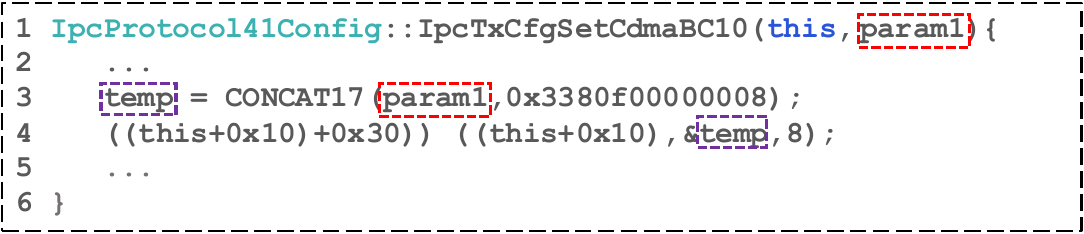}
    \caption{Example of a direct input parameter for configuring the baseband.}
    \label{fig:hybrid1}
\end{figure}

\begin{figure}[t]
    \centering
    \includegraphics[width=0.4\textwidth]{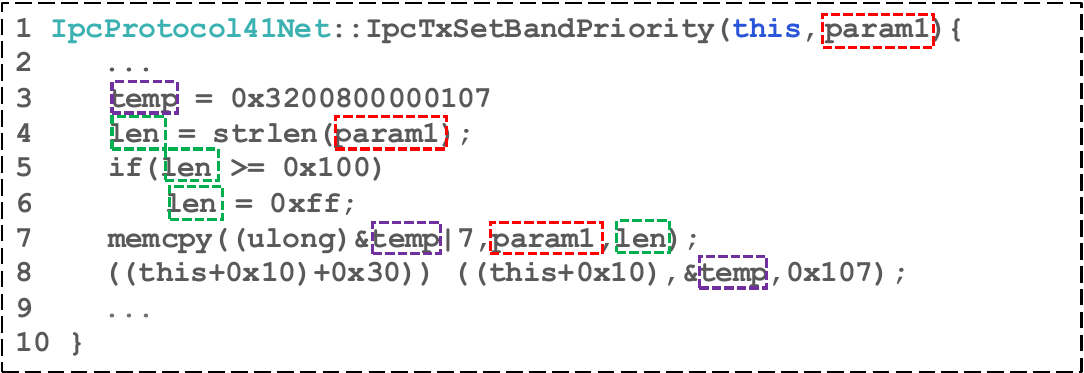}
    \caption{Example of a derived parameter setting priority.}
    \label{fig:hybrid2}
\end{figure}

\begin{figure}[t]
    \centering
    \includegraphics[width=0.4\textwidth]{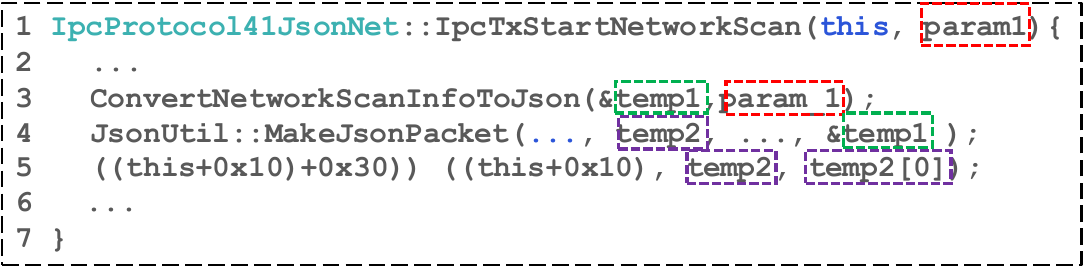}
    \caption{Example of a structured parameter of JSON format for starting network scan.}
    \label{fig:hybrid3}
\end{figure}

\paragraph{Direct Input Parameter}
\autoref{fig:hybrid1} shows a baseband command for the RIL to adjust the CDMA configurations in the baseband. The command payload is constructed by concatenating a byte variable (\texttt{param1}) from the function parameter with a constant hex payload. This hybrid command depends on a variable \texttt{param1} that is essentially an input from external sources such as a user-space application. It can also be inferred from the payload size (8 bytes in total) that the parameter occupies one single byte in total, possibly indicating a boolean variable.

\paragraph{Derived Parameter}
Similar to the previous example, the baseband command shown in \autoref{fig:hybrid2} is also initialized with a constant hex payload prefix. However, it depends on a parameter of string type, as indicated in line 4 since the parameter is taken by a \texttt{strlen} function to obtain its length. Next, the function performs a truncation that limits the parameter size to 255 (i.e., \texttt{0xff}) at maximum. The parameter string is to indicate the band priority. Without prior information, e.g., a valid seed input, it is thus hard for a fuzzer to generate feasible test cases for this command payload.

\paragraph{Structured Parameter}
\autoref{fig:hybrid3} presents an example of a command parameter constructed as a complex data structure. Based on the function semantics, the command is to instruct the baseband the start network scanning. To this end, the function first converts the network scanning information to a JSON object, by using the provided input \texttt{param1}, and then further builds the command payload as a JSON packet. For simplicity, we do not show the actual implementation of the two utility functions, as their functionality is already suggested by the function names.

%% file: table/diff.tex
\begin{table}[t]
    \centering
     \scriptsize
    \begin{tabular}{lrrrrrr}
    \toprule
        \multirow{2}{*}{\textbf{Model}} & \multicolumn{3}{c}{\textbf{Write}} & \multicolumn{3}{c}{\textbf{Read}} \\ 
        \cmidrule(lr){2-4} \cmidrule(lr){5-7}
        ~ & Cmds & BaseUni & CurUni & Cmds & BaseUni & CurUni \\
        \midrule
        N975 & 478 & / & / & 49 & / & / \\ 
        J600 & 427 & 63 & 12 & 53 & 0 & 5 \\ 
        G920 & 570 & 87 & 179 & 67 & 5 & 22 \\ 
        G930 & 433 & 67 & 22 & 53 & 0 & 5 \\ 
        G960 & 493 & 10 & 25 & 55 & 0 & 6 \\ 
        G950 & 445 & 48 & 15 & 53 & 0 & 5 \\ 
        N950 & 429 & 63 & 14 & 53 & 0 & 5 \\ 
        N960 & 493 & 10 & 25 & 55 & 0 & 6 \\
        \midrule
        A205 & 505 & 22 & 1 & 204 & 5 & 20 \\ 
        A305 & 505 & 22 & 1 & 204 & 5 & 20 \\ 
        A405 & 505 & 22 & 1 & 204 & 5 & 20 \\ 
        A415 & 505 & 22 & 1 & 204 & 5 & 20 \\ 
        A505 & 505 & 22 & 1 & 204 & 5 & 20 \\ 
        A606 & 505 & 22 & 1 & 204 & 5 & 20 \\ 
        A705 & 505 & 22 & 1 & 204 & 5 & 20 \\ 
        A805 & 505 & 22 & 1 & 204 & 5 & 20 \\ 
        G981 & 525 & 1 & 0 & 205 & 2 & 18 \\ 
        G991 & 525 & 1 & 0 & 205 & 2 & 18 \\ 
        N981 & 525 & 1 & 0 & 205 & 2 & 18 \\ 
        A426 & 525 & 1 & 0 & 204 & 2 & 17 \\ 
        A526 & 525 & 1 & 0 & 204 & 2 & 17 \\ 
        A725 & 525 & 1 & 0 & 204 & 2 & 17 \\ 
        A908 & 525 & 1 & 0 & 204 & 2 & 17 \\ 
        A515 & 526 & 0 & 0 & 189 & 0 & 0 \\ 
        A516 & 526 & 0 & 0 & 189 & 0 & 0 \\ 
        G973 & 526 & 0 & 0 & 189 & 0 & 0 \\ 
        G977 & 526 & 0 & 0 & 189 & 0 & 0 \\ 
        A536 & 526 & / & / & 189 & / & / \\ \bottomrule
    \end{tabular}
    \caption{Discrepancy in Write and Read related commands between each firmware and the Baseline resident in \texttt{Ipc41} (Upper) and \texttt{Ipc41X} (Lower) separately.}
    \label{tab:diff}
\end{table}






%% file: paper.bbl

\begin{thebibliography}{66}


\ifx \showCODEN    \undefined \def \showCODEN     #1{\unskip}     \fi
\ifx \showDOI      \undefined \def \showDOI       #1{#1}\fi
\ifx \showISBNx    \undefined \def \showISBNx     #1{\unskip}     \fi
\ifx \showISBNxiii \undefined \def \showISBNxiii  #1{\unskip}     \fi
\ifx \showISSN     \undefined \def \showISSN      #1{\unskip}     \fi
\ifx \showLCCN     \undefined \def \showLCCN      #1{\unskip}     \fi
\ifx \shownote     \undefined \def \shownote      #1{#1}          \fi
\ifx \showarticletitle \undefined \def \showarticletitle #1{#1}   \fi
\ifx \showURL      \undefined \def \showURL       {\relax}        \fi
\providecommand\bibfield[2]{#2}
\providecommand\bibinfo[2]{#2}
\providecommand\natexlab[1]{#1}
\providecommand\showeprint[2][]{arXiv:#2}

\bibitem[Allen(1970)]%
        {allen1970control}
\bibfield{author}{\bibinfo{person}{Frances~E Allen}.} \bibinfo{year}{1970}\natexlab{}.
\newblock \showarticletitle{Control flow analysis}. In \bibinfo{booktitle}{\emph{ACM Sigplan Notices}}.
\newblock


\bibitem[{Android}(2022)]%
        {android_ril}
\bibfield{author}{\bibinfo{person}{{Android}}.} \bibinfo{year}{2022}\natexlab{}.
\newblock \bibinfo{title}{RIL Refactoring | Android Open Source Project}.
\newblock \bibinfo{howpublished}{\url{https://source.android.com/devices/tech/connect/ril}}.
\newblock


\bibitem[{Android}(2024a)]%
        {android_ipc}
\bibfield{author}{\bibinfo{person}{{Android}}.} \bibinfo{year}{2024}\natexlab{a}.
\newblock \bibinfo{title}{Android Interface Definition Language (AIDL)}.
\newblock \bibinfo{howpublished}{\url{https://developer.android.com/guide/components/aidl}}.
\newblock


\bibitem[{Android}(2024b)]%
        {aosp}
\bibfield{author}{\bibinfo{person}{{Android}}.} \bibinfo{year}{2024}\natexlab{b}.
\newblock \bibinfo{title}{Android Open Source Project}.
\newblock \bibinfo{howpublished}{\url{https://source.android.com/}}.
\newblock


\bibitem[Bacon and Sweeney(1996)]%
        {bacon1996fast}
\bibfield{author}{\bibinfo{person}{David~F Bacon} {and} \bibinfo{person}{Peter~F Sweeney}.} \bibinfo{year}{1996}\natexlab{}.
\newblock \showarticletitle{Fast static analysis of C++ virtual function calls}. In \bibinfo{booktitle}{\emph{11th ACM SIGPLAN conference on Object-oriented programming, systems, languages, and applications (OOPSLA)}}.
\newblock


\bibitem[Berard and Fargues(2020)]%
        {berard2020design}
\bibfield{author}{\bibinfo{person}{David Berard} {and} \bibinfo{person}{Vincent Fargues}.} \bibinfo{year}{2020}\natexlab{}.
\newblock \showarticletitle{How to design a baseband debugger}. In \bibinfo{booktitle}{\emph{Information and Communication Technology Security Symposium (SSTIC)}}.
\newblock


\bibitem[Bl{\'a}zquez et~al\mbox{.}(2021)]%
        {blazquez2021trouble}
\bibfield{author}{\bibinfo{person}{Eduardo Bl{\'a}zquez}, \bibinfo{person}{Sergio Pastrana}, \bibinfo{person}{{\'A}lvaro Feal}, \bibinfo{person}{Julien Gamba}, \bibinfo{person}{Platon Kotzias}, \bibinfo{person}{Narseo Vallina-Rodriguez}, {and} \bibinfo{person}{Juan Tapiador}.} \bibinfo{year}{2021}\natexlab{}.
\newblock \showarticletitle{Trouble over-the-air: An analysis of fota apps in the android ecosystem}. In \bibinfo{booktitle}{\emph{42nd IEEE Symposium on Security and Privacy (SP)}}.
\newblock


\bibitem[B{\"o}hme et~al\mbox{.}(2017)]%
        {bohme2017directed}
\bibfield{author}{\bibinfo{person}{Marcel B{\"o}hme}, \bibinfo{person}{Van-Thuan Pham}, \bibinfo{person}{Manh-Dung Nguyen}, {and} \bibinfo{person}{Abhik Roychoudhury}.} \bibinfo{year}{2017}\natexlab{}.
\newblock \showarticletitle{Directed greybox fuzzing}. In \bibinfo{booktitle}{\emph{24th ACM SIGSAC Conference on Computer and Communications Security (CCS)}}.
\newblock


\bibitem[Carikli(2021)]%
        {samsung_ipc}
\bibfield{author}{\bibinfo{person}{Denis~'GNUtoo' Carikli}.} \bibinfo{year}{2021}\natexlab{}.
\newblock \bibinfo{title}{Samsung-ipc}.
\newblock \bibinfo{howpublished}{\url{https://redmine.replicant.us/projects/replicant/wiki/Samsung-ipc}}.
\newblock


\bibitem[El-Rewini and Aafer(2021)]%
        {el2021dissecting}
\bibfield{author}{\bibinfo{person}{Zeinab El-Rewini} {and} \bibinfo{person}{Yousra Aafer}.} \bibinfo{year}{2021}\natexlab{}.
\newblock \showarticletitle{Dissecting residual APIs in custom android ROMs}. In \bibinfo{booktitle}{\emph{28th ACM SIGSAC Conference on Computer and Communications Security (CCS)}}.
\newblock


\bibitem[Elsabagh et~al\mbox{.}(2017)]%
        {elsabagh2017strict}
\bibfield{author}{\bibinfo{person}{Mohamed Elsabagh}, \bibinfo{person}{Dan Fleck}, {and} \bibinfo{person}{Angelos Stavrou}.} \bibinfo{year}{2017}\natexlab{}.
\newblock \showarticletitle{Strict virtual call integrity checking for C++ binaries}. In \bibinfo{booktitle}{\emph{17th ACM on Asia Conference on Computer and Communications Security (ASIACCS)}}.
\newblock


\bibitem[Elsabagh et~al\mbox{.}(2020)]%
        {elsabagh2020firmscope}
\bibfield{author}{\bibinfo{person}{Mohamed Elsabagh}, \bibinfo{person}{Ryan Johnson}, \bibinfo{person}{Angelos Stavrou}, \bibinfo{person}{Chaoshun Zuo}, \bibinfo{person}{Qingchuan Zhao}, {and} \bibinfo{person}{Zhiqiang Lin}.} \bibinfo{year}{2020}\natexlab{}.
\newblock \showarticletitle{FIRMSCOPE: Automatic uncovering of Privilege-Escalation vulnerabilities in Pre-Installed apps in android firmware}. In \bibinfo{booktitle}{\emph{29th USENIX Security Symposium (USENIX Security)}}.
\newblock


\bibitem[Erinfolami and Prakash(2019)]%
        {erinfolami2019declassifier}
\bibfield{author}{\bibinfo{person}{Rukayat~Ayomide Erinfolami} {and} \bibinfo{person}{Aravind Prakash}.} \bibinfo{year}{2019}\natexlab{}.
\newblock \showarticletitle{DeClassifier: Class-Inheritance Inference Engine for Optimized C++ Binaries}. In \bibinfo{booktitle}{\emph{19th ACM Asia Conference on Computer and Communications Security (ASIACCS)}}.
\newblock


\bibitem[Erinfolami and Prakash(2020)]%
        {erinfolami2020devil}
\bibfield{author}{\bibinfo{person}{Rukayat~Ayomide Erinfolami} {and} \bibinfo{person}{Aravind Prakash}.} \bibinfo{year}{2020}\natexlab{}.
\newblock \showarticletitle{Devil is virtual: Reversing virtual inheritance in C++ binaries}. In \bibinfo{booktitle}{\emph{27th ACM SIGSAC Conference on Computer and Communications Security (CCS)}}.
\newblock


\bibitem[Fioraldi et~al\mbox{.}(2020)]%
        {fioraldi2020afl++}
\bibfield{author}{\bibinfo{person}{Andrea Fioraldi}, \bibinfo{person}{Dominik Maier}, \bibinfo{person}{Heiko Ei{\ss}feldt}, {and} \bibinfo{person}{Marc Heuse}.} \bibinfo{year}{2020}\natexlab{}.
\newblock \showarticletitle{AFL++: Combining incremental steps of fuzzing research}. In \bibinfo{booktitle}{\emph{14th USENIX Conference on Offensive Technologies (WOOT)}}.
\newblock


\bibitem[Fokin et~al\mbox{.}(2011)]%
        {fokin2011smartdec}
\bibfield{author}{\bibinfo{person}{Alexander Fokin}, \bibinfo{person}{Egor Derevenetc}, \bibinfo{person}{Alexander Chernov}, {and} \bibinfo{person}{Katerina Troshina}.} \bibinfo{year}{2011}\natexlab{}.
\newblock \showarticletitle{SmartDec: approaching C++ decompilation}. In \bibinfo{booktitle}{\emph{18th Working Conference on Reverse Engineering (WCRE)}}.
\newblock


\bibitem[Ghidra(2024)]%
        {pcode}
\bibfield{author}{\bibinfo{person}{Ghidra}.} \bibinfo{year}{2024}\natexlab{}.
\newblock \bibinfo{title}{P-Code}.
\newblock \bibinfo{howpublished}{\url{https://ghidra.re/ghidra_docs/api/ghidra/program/model/pcode/package-summary.html}}.
\newblock


\bibitem[Golde and Komaromy(2016)]%
        {golde2016breaking}
\bibfield{author}{\bibinfo{person}{Nico Golde} {and} \bibinfo{person}{Daniel Komaromy}.} \bibinfo{year}{2016}\natexlab{}.
\newblock \showarticletitle{Breaking Band: reverse engineering and exploiting the shannon baseband}. In \bibinfo{booktitle}{\emph{2016 Recon}}.
\newblock


\bibitem[Grassi and Chen(2021)]%
        {grassi2021over}
\bibfield{author}{\bibinfo{person}{Marco Grassi} {and} \bibinfo{person}{Xingyu Chen}.} \bibinfo{year}{2021}\natexlab{}.
\newblock \showarticletitle{Over The Air Baseband Exploit: Gaining Remote Code Execution on 5G Smartphones}. In \bibinfo{booktitle}{\emph{BlackHat USA}}.
\newblock


\bibitem[Grassi et~al\mbox{.}(2018)]%
        {grassi2018exploitation}
\bibfield{author}{\bibinfo{person}{Marco Grassi}, \bibinfo{person}{Muqing Liu}, {and} \bibinfo{person}{Tianyi Xie}.} \bibinfo{year}{2018}\natexlab{}.
\newblock \showarticletitle{Exploitation Of A Modern Smartphone Baseband}. In \bibinfo{booktitle}{\emph{BlackHat USA}}.
\newblock


\bibitem[Hazarika(2023)]%
        {android_hidden_code}
\bibfield{author}{\bibinfo{person}{Skanda Hazarika}.} \bibinfo{year}{2023}\natexlab{}.
\newblock \bibinfo{title}{Android hidden codes: All the custom dialer codes and what they do}.
\newblock \bibinfo{howpublished}{\url{https://www.xda-developers.com/android-secret-codes/}}.
\newblock


\bibitem[Hernandez and Butler(2019)]%
        {hernandez2019basebads}
\bibfield{author}{\bibinfo{person}{Grant Hernandez} {and} \bibinfo{person}{Kevin~RB Butler}.} \bibinfo{year}{2019}\natexlab{}.
\newblock \showarticletitle{Basebads: Automated security analysis of baseband firmware}. In \bibinfo{booktitle}{\emph{12th Conference on Security and Privacy in Wireless and Mobile Networks (WiSec)}}.
\newblock


\bibitem[Hernandez et~al\mbox{.}(2022)]%
        {hernandez_firmwire_2022}
\bibfield{author}{\bibinfo{person}{Grant Hernandez}, \bibinfo{person}{Marius Muench}, \bibinfo{person}{Dominik Maier}, \bibinfo{person}{Alyssa Milburn}, \bibinfo{person}{Shinjo Park}, \bibinfo{person}{Tobias Scharnowski}, \bibinfo{person}{Tyler Tucker}, \bibinfo{person}{Patrick Traynor}, {and} \bibinfo{person}{Kevin R.~B. Butler}.} \bibinfo{year}{2022}\natexlab{}.
\newblock \showarticletitle{{FirmWire: Transparent Dynamic Analysis for Cellular Baseband Firmware}}. In \bibinfo{booktitle}{\emph{{29th Network and Distributed System Security Symposium (NDSS)}}}.
\newblock


\bibitem[Hernandez et~al\mbox{.}(2020)]%
        {hernandez2020bigmac}
\bibfield{author}{\bibinfo{person}{Grant Hernandez}, \bibinfo{person}{Dave~Jing Tian}, \bibinfo{person}{Anurag~Swarnim Yadav}, \bibinfo{person}{Byron~J Williams}, {and} \bibinfo{person}{Kevin~RB Butler}.} \bibinfo{year}{2020}\natexlab{}.
\newblock \showarticletitle{BigMAC:Fine-Grained Policy Analysis of Android Firmware}. In \bibinfo{booktitle}{\emph{29th USENIX Security Symposium (USENIX Security)}}.
\newblock


\bibitem[Hou et~al\mbox{.}(2022)]%
        {hou2022large}
\bibfield{author}{\bibinfo{person}{Qinsheng Hou}, \bibinfo{person}{Wenrui Diao}, \bibinfo{person}{Yanhao Wang}, \bibinfo{person}{Xiaofeng Liu}, \bibinfo{person}{Song Liu}, \bibinfo{person}{Lingyun Ying}, \bibinfo{person}{Shanqing Guo}, \bibinfo{person}{Yuanzhi Li}, \bibinfo{person}{Meining Nie}, {and} \bibinfo{person}{Haixin Duan}.} \bibinfo{year}{2022}\natexlab{}.
\newblock \showarticletitle{Large-scale security measurements on the android firmware ecosystem}. In \bibinfo{booktitle}{\emph{44th International Conference on Software Engineering (ICSE)}}.
\newblock


\bibitem[Hussain et~al\mbox{.}(2021)]%
        {hussain2021noncompliance}
\bibfield{author}{\bibinfo{person}{Syed~Rafiul Hussain}, \bibinfo{person}{Imtiaz Karim}, \bibinfo{person}{Abdullah~Al Ishtiaq}, \bibinfo{person}{Omar Chowdhury}, {and} \bibinfo{person}{Elisa Bertino}.} \bibinfo{year}{2021}\natexlab{}.
\newblock \showarticletitle{Noncompliance as Deviant Behavior: An Automated Black-box Noncompliance Checker for 4G LTE Cellular Devices}. In \bibinfo{booktitle}{\emph{28th ACM SIGSAC Conference on Computer and Communications Security (CCS)}}.
\newblock


\bibitem[Jin et~al\mbox{.}(2014)]%
        {jin2014recovering}
\bibfield{author}{\bibinfo{person}{Wesley Jin}, \bibinfo{person}{Cory Cohen}, \bibinfo{person}{Jeffrey Gennari}, \bibinfo{person}{Charles Hines}, \bibinfo{person}{Sagar Chaki}, \bibinfo{person}{Arie Gurfinkel}, \bibinfo{person}{Jeffrey Havrilla}, {and} \bibinfo{person}{Priya Narasimhan}.} \bibinfo{year}{2014}\natexlab{}.
\newblock \showarticletitle{Recovering C++ objects from binaries using inter-procedural data-flow analysis}. In \bibinfo{booktitle}{\emph{4th ACM SIGPLAN on Program Protection and Reverse Engineering Workshop (PPREW)}}.
\newblock


\bibitem[K(2014)]%
        {samsung_backdoor}
\bibfield{author}{\bibinfo{person}{Paul K}.} \bibinfo{year}{2014}\natexlab{}.
\newblock \bibinfo{title}{Replicant developers find and close Samsung Galaxy backdoor}.
\newblock \bibinfo{howpublished}{\url{https://www.fsf.org/blogs/community/replicant-developers-find-and-close-samsung-galaxy-backdoor}}.
\newblock


\bibitem[Karim et~al\mbox{.}(2019)]%
        {karim2019opening}
\bibfield{author}{\bibinfo{person}{Imtiaz Karim}, \bibinfo{person}{Fabrizio Cicala}, \bibinfo{person}{Syed~Rafiul Hussain}, \bibinfo{person}{Omar Chowdhury}, {and} \bibinfo{person}{Elisa Bertino}.} \bibinfo{year}{2019}\natexlab{}.
\newblock \showarticletitle{Opening Pandora's box through ATFuzzer: dynamic analysis of AT interface for Android smartphones}. In \bibinfo{booktitle}{\emph{35th Annual Computer Security Applications Conference (ACSAC)}}.
\newblock


\bibitem[Kim et~al\mbox{.}(2023)]%
        {kim2023basecomp}
\bibfield{author}{\bibinfo{person}{Eunsoo Kim}, \bibinfo{person}{Min~Woo Baek}, \bibinfo{person}{CheolJun Park}, \bibinfo{person}{Dongkwan Kim}, \bibinfo{person}{Yongdae Kim}, {and} \bibinfo{person}{Insu Yun}.} \bibinfo{year}{2023}\natexlab{}.
\newblock \showarticletitle{BASECOMP: A Comparative Analysis for Integrity Protection in Cellular Baseband Software}. In \bibinfo{booktitle}{\emph{32nd USENIX Security Symposium (USENIX Security)}}.
\newblock


\bibitem[Kim et~al\mbox{.}(2021)]%
        {kim2021basespec}
\bibfield{author}{\bibinfo{person}{Eunsoo Kim}, \bibinfo{person}{Dongkwan Kim}, \bibinfo{person}{CheolJun Park}, \bibinfo{person}{Insu Yun}, {and} \bibinfo{person}{Yongdae Kim}.} \bibinfo{year}{2021}\natexlab{}.
\newblock \showarticletitle{BASESPEC: Comparative Analysis of Baseband Software and Cellular Specifications for L3 Protocols}. In \bibinfo{booktitle}{\emph{28th Network and Distributed System Security Symposium (NDSS)}}.
\newblock


\bibitem[Kr{\"o}ll et~al\mbox{.}(2021)]%
        {kroll2021aristoteles}
\bibfield{author}{\bibinfo{person}{Tobias Kr{\"o}ll}, \bibinfo{person}{Stephan Kleber}, \bibinfo{person}{Frank Kargl}, \bibinfo{person}{Matthias Hollick}, {and} \bibinfo{person}{Jiska Classen}.} \bibinfo{year}{2021}\natexlab{}.
\newblock \showarticletitle{ARIstoteles--Dissecting Apple’s Baseband Interface}. In \bibinfo{booktitle}{\emph{26th European Symposium on Research in Computer Security (ESORICS)}}.
\newblock


\bibitem[laforge(2019)]%
        {qualcomm_ipc}
\bibfield{author}{\bibinfo{person}{laforge}.} \bibinfo{year}{2019}\natexlab{}.
\newblock \bibinfo{title}{Qualcomm Linux Modems by Quectel \& Co}.
\newblock \bibinfo{howpublished}{\url{https://osmocom.org/projects/quectel-modems/wiki/QMI}}.
\newblock


\bibitem[Li et~al\mbox{.}(2021)]%
        {li2021para}
\bibfield{author}{\bibinfo{person}{Wenqiang Li}, \bibinfo{person}{Le Guan}, \bibinfo{person}{Jingqiang Lin}, \bibinfo{person}{Jiameng Shi}, {and} \bibinfo{person}{Fengjun Li}.} \bibinfo{year}{2021}\natexlab{}.
\newblock \showarticletitle{From Library Portability to Para-rehosting: Natively Executing Microcontroller Software on Commodity Hardware}. In \bibinfo{booktitle}{\emph{28th Network and Distributed System Security Symposium (NDSS)}}.
\newblock


\bibitem[Li et~al\mbox{.}(2022)]%
        {li2022muafl}
\bibfield{author}{\bibinfo{person}{Wenqiang Li}, \bibinfo{person}{Jiameng Shi}, \bibinfo{person}{Fengjun Li}, \bibinfo{person}{Jingqiang Lin}, \bibinfo{person}{Wei Wang}, {and} \bibinfo{person}{Le Guan}.} \bibinfo{year}{2022}\natexlab{}.
\newblock \showarticletitle{$\mu$AFL: non-intrusive feedback-driven fuzzing for microcontroller firmware}. In \bibinfo{booktitle}{\emph{44th International Conference on Software Engineering (ICSE)}}.
\newblock


\bibitem[Li et~al\mbox{.}(2016)]%
        {li2016mobileinsight}
\bibfield{author}{\bibinfo{person}{Yuanjie Li}, \bibinfo{person}{Chunyi Peng}, \bibinfo{person}{Zengwen Yuan}, \bibinfo{person}{Jiayao Li}, \bibinfo{person}{Haotian Deng}, {and} \bibinfo{person}{Tao Wang}.} \bibinfo{year}{2016}\natexlab{}.
\newblock \showarticletitle{Mobileinsight: Extracting and analyzing cellular network information on smartphones}. In \bibinfo{booktitle}{\emph{22nd Annual International Conference on Mobile Computing and Networking (MobiCom)}}.
\newblock


\bibitem[Liu et~al\mbox{.}(2024)]%
        {liu2024semantic}
\bibfield{author}{\bibinfo{person}{Yiming Liu}, \bibinfo{person}{Cen Zhang}, \bibinfo{person}{Feng Li}, \bibinfo{person}{Yeting Li}, \bibinfo{person}{Jianhua Zhou}, \bibinfo{person}{Jian Wang}, \bibinfo{person}{Lanlan Zhan}, \bibinfo{person}{Yang Liu}, {and} \bibinfo{person}{Wei Huo}.} \bibinfo{year}{2024}\natexlab{}.
\newblock \showarticletitle{Semantic-Enhanced Static Vulnerability Detection in Baseband Firmware}. In \bibinfo{booktitle}{\emph{46th International Conference on Software Engineering (ICSE)}}.
\newblock


\bibitem[Lu and Hu(2019)]%
        {lu2019does}
\bibfield{author}{\bibinfo{person}{Kangjie Lu} {and} \bibinfo{person}{Hong Hu}.} \bibinfo{year}{2019}\natexlab{}.
\newblock \showarticletitle{Where does it go? refining indirect-call targets with multi-layer type analysis}. In \bibinfo{booktitle}{\emph{26th ACM SIGSAC Conference on Computer and Communications Security (CCS)}}.
\newblock


\bibitem[Maier et~al\mbox{.}(2020)]%
        {maier2020basesafe}
\bibfield{author}{\bibinfo{person}{Dominik Maier}, \bibinfo{person}{Lukas Seidel}, {and} \bibinfo{person}{Shinjo Park}.} \bibinfo{year}{2020}\natexlab{}.
\newblock \showarticletitle{Basesafe: Baseband sanitized fuzzing through emulation}. In \bibinfo{booktitle}{\emph{13th ACM Conference on Security and Privacy in Wireless and Mobile Networks (WiSec)}}.
\newblock


\bibitem[Ming et~al\mbox{.}(2012)]%
        {ming2012ibinhunt}
\bibfield{author}{\bibinfo{person}{Jiang Ming}, \bibinfo{person}{Meng Pan}, {and} \bibinfo{person}{Debin Gao}.} \bibinfo{year}{2012}\natexlab{}.
\newblock \showarticletitle{iBinHunt: Binary hunting with inter-procedural control flow}. In \bibinfo{booktitle}{\emph{9th International Conference on Information Security and Cryptology (ICISC)}}.
\newblock


\bibitem[NSA(2024)]%
        {ghidra}
\bibfield{author}{\bibinfo{person}{NSA}.} \bibinfo{year}{2024}\natexlab{}.
\newblock \bibinfo{title}{Ghidra}.
\newblock \bibinfo{howpublished}{\url{https://ghidra-sre.org/}}.
\newblock


\bibitem[Paleari(2024)]%
        {samsung_rild}
\bibfield{author}{\bibinfo{person}{Roberto Paleari}.} \bibinfo{year}{2024}\natexlab{}.
\newblock \bibinfo{title}{Interacting with Samsung radio layer (RILD)}.
\newblock \bibinfo{howpublished}{\url{http://roberto.greyhats.it/2016/05/samsung-access-rild.html}}.
\newblock


\bibitem[Pawlowski et~al\mbox{.}(2017)]%
        {pawlowski2017marx}
\bibfield{author}{\bibinfo{person}{Andre Pawlowski}, \bibinfo{person}{Moritz Contag}, \bibinfo{person}{Victor van~der Veen}, \bibinfo{person}{Chris Ouwehand}, \bibinfo{person}{Thorsten Holz}, \bibinfo{person}{Herbert Bos}, \bibinfo{person}{Elias Athanasopoulos}, {and} \bibinfo{person}{Cristiano Giuffrida}.} \bibinfo{year}{2017}\natexlab{}.
\newblock \showarticletitle{MARX: Uncovering Class Hierarchies in C++ Programs.}. In \bibinfo{booktitle}{\emph{24th Network and Distributed System Security Symposium (NDSS)}}.
\newblock


\bibitem[Pawlowski et~al\mbox{.}(2019)]%
        {pawlowski2019vps}
\bibfield{author}{\bibinfo{person}{Andre Pawlowski}, \bibinfo{person}{Victor van~der Veen}, \bibinfo{person}{Dennis Andriesse}, \bibinfo{person}{Erik van~der Kouwe}, \bibinfo{person}{Thorsten Holz}, \bibinfo{person}{Cristiano Giuffrida}, {and} \bibinfo{person}{Herbert Bos}.} \bibinfo{year}{2019}\natexlab{}.
\newblock \showarticletitle{VPS: excavating high-level C++ constructs from low-level binaries to protect dynamic dispatching}. In \bibinfo{booktitle}{\emph{35th Annual Computer Security Applications Conference (ACSAC)}}.
\newblock


\bibitem[Possemato et~al\mbox{.}(2021)]%
        {possemato2021trust}
\bibfield{author}{\bibinfo{person}{Andrea Possemato}, \bibinfo{person}{Simone Aonzo}, \bibinfo{person}{Davide Balzarotti}, {and} \bibinfo{person}{Yanick Fratantonio}.} \bibinfo{year}{2021}\natexlab{}.
\newblock \showarticletitle{Trust, but verify: A longitudinal analysis of Android OEM compliance and customization}. In \bibinfo{booktitle}{\emph{42nd IEEE Symposium on Security and Privacy (SP)}}.
\newblock


\bibitem[Ravnås(2024)]%
        {frida}
\bibfield{author}{\bibinfo{person}{Ole André~Vadla Ravnås}.} \bibinfo{year}{2024}\natexlab{}.
\newblock \bibinfo{title}{Frida - A world-class dynamic instrumentation toolkit}.
\newblock \bibinfo{howpublished}{\url{https://frida.re/}}.
\newblock


\bibitem[Rays(2024)]%
        {ida}
\bibfield{author}{\bibinfo{person}{Hex Rays}.} \bibinfo{year}{2024}\natexlab{}.
\newblock \bibinfo{title}{IDA Pro}.
\newblock \bibinfo{howpublished}{\url{https://www.hex-rays.com/idapro}}.
\newblock


\bibitem[SAMMOBILE(2024)]%
        {SamMobil36:online}
\bibfield{author}{\bibinfo{person}{SAMMOBILE}.} \bibinfo{year}{2024}\natexlab{}.
\newblock \bibinfo{title}{SamMobile - Your source for all Samsung news}.
\newblock \bibinfo{howpublished}{\url{https://www.sammobile.com/}}.
\newblock


\bibitem[Schubert et~al\mbox{.}(2019)]%
        {schubert2019phasar}
\bibfield{author}{\bibinfo{person}{Philipp~Dominik Schubert}, \bibinfo{person}{Ben Hermann}, {and} \bibinfo{person}{Eric Bodden}.} \bibinfo{year}{2019}\natexlab{}.
\newblock \showarticletitle{Phasar: An inter-procedural static analysis framework for C/C++}. In \bibinfo{booktitle}{\emph{25th International Conference on Tools and Algorithms for the Construction and Analysis of Systems (TACAS)}}.
\newblock


\bibitem[Schwartz et~al\mbox{.}(2018)]%
        {schwartz2018using}
\bibfield{author}{\bibinfo{person}{Edward~J Schwartz}, \bibinfo{person}{Cory~F Cohen}, \bibinfo{person}{Michael Duggan}, \bibinfo{person}{Jeffrey Gennari}, \bibinfo{person}{Jeffrey~S Havrilla}, {and} \bibinfo{person}{Charles Hines}.} \bibinfo{year}{2018}\natexlab{}.
\newblock \showarticletitle{Using logic programming to recover C++ classes and methods from compiled executables}. In \bibinfo{booktitle}{\emph{25th ACM SIGSAC Conference on Computer and Communications Security (CCS)}}.
\newblock


\bibitem[Shao et~al\mbox{.}(2016)]%
        {shao2016kratos}
\bibfield{author}{\bibinfo{person}{Yuru Shao}, \bibinfo{person}{Qi~Alfred Chen}, \bibinfo{person}{Zhuoqing~Morley Mao}, \bibinfo{person}{Jason Ott}, {and} \bibinfo{person}{Zhiyun Qian}.} \bibinfo{year}{2016}\natexlab{}.
\newblock \showarticletitle{Kratos: Discovering Inconsistent Security Policy Enforcement in the Android Framework}. In \bibinfo{booktitle}{\emph{23th Network and Distributed System Security Symposium (NDSS)}}.
\newblock


\bibitem[Shi et~al\mbox{.}(2024)]%
        {shi2024invivo}
\bibfield{author}{\bibinfo{person}{Jiameng Shi}, \bibinfo{person}{Wenqiang Li}, \bibinfo{person}{Wenwen Wang}, {and} \bibinfo{person}{Le Guan}.} \bibinfo{year}{2024}\natexlab{}.
\newblock \showarticletitle{Facilitating Non-Intrusive In-Vivo Firmware Testing with Stateless Instrumentation}. In \bibinfo{booktitle}{\emph{31st Network and Distributed System Security Symposium (NDSS)}}.
\newblock


\bibitem[Shoshitaishvili et~al\mbox{.}(2016)]%
        {shoshitaishvili2016sok}
\bibfield{author}{\bibinfo{person}{Yan Shoshitaishvili}, \bibinfo{person}{Ruoyu Wang}, \bibinfo{person}{Christopher Salls}, \bibinfo{person}{Nick Stephens}, \bibinfo{person}{Mario Polino}, \bibinfo{person}{Andrew Dutcher}, \bibinfo{person}{John Grosen}, \bibinfo{person}{Siji Feng}, \bibinfo{person}{Christophe Hauser}, \bibinfo{person}{Christopher Kruegel}, {et~al\mbox{.}}} \bibinfo{year}{2016}\natexlab{}.
\newblock \showarticletitle{SOK:(State of) The Art of War: Offensive techniques in binary analysis}. In \bibinfo{booktitle}{\emph{37th IEEE Symposium on Security and Privacy (SP)}}.
\newblock


\bibitem[Tian et~al\mbox{.}(2018)]%
        {tian2018attention}
\bibfield{author}{\bibinfo{person}{Dave~Jing Tian}, \bibinfo{person}{Grant Hernandez}, \bibinfo{person}{Joseph~I Choi}, \bibinfo{person}{Vanessa Frost}, \bibinfo{person}{Christie Raules}, \bibinfo{person}{Patrick Traynor}, \bibinfo{person}{Hayawardh Vijayakumar}, \bibinfo{person}{Lee Harrison}, \bibinfo{person}{Amir Rahmati}, \bibinfo{person}{Michael Grace}, {et~al\mbox{.}}} \bibinfo{year}{2018}\natexlab{}.
\newblock \showarticletitle{Attention spanned: Comprehensive vulnerability analysis of AT commands within the android ecosystem}. In \bibinfo{booktitle}{\emph{27th USENIX Security Symposium (USENIX Security)}}.
\newblock


\bibitem[Vallina-Rodriguez et~al\mbox{.}(2013)]%
        {vallina2013rilanalyzer}
\bibfield{author}{\bibinfo{person}{Narseo Vallina-Rodriguez}, \bibinfo{person}{Andrius Au{\c{c}}inas}, \bibinfo{person}{Mario Almeida}, \bibinfo{person}{Yan Grunenberger}, \bibinfo{person}{Konstantina Papagiannaki}, {and} \bibinfo{person}{Jon Crowcroft}.} \bibinfo{year}{2013}\natexlab{}.
\newblock \showarticletitle{RILAnalyzer: a comprehensive 3G monitor on your phone}. In \bibinfo{booktitle}{\emph{13th Internet Measurement Conference (IMC)}}.
\newblock


\bibitem[Van Der~Veen et~al\mbox{.}(2016)]%
        {van2016tough}
\bibfield{author}{\bibinfo{person}{Victor Van Der~Veen}, \bibinfo{person}{Enes G{\"o}ktas}, \bibinfo{person}{Moritz Contag}, \bibinfo{person}{Andre Pawoloski}, \bibinfo{person}{Xi Chen}, \bibinfo{person}{Sanjay Rawat}, \bibinfo{person}{Herbert Bos}, \bibinfo{person}{Thorsten Holz}, \bibinfo{person}{Elias Athanasopoulos}, {and} \bibinfo{person}{Cristiano Giuffrida}.} \bibinfo{year}{2016}\natexlab{}.
\newblock \showarticletitle{A tough call: Mitigating advanced code-reuse attacks at the binary level}. In \bibinfo{booktitle}{\emph{37th IEEE Symposium on Security and Privacy (SP)}}.
\newblock


\bibitem[Wang et~al\mbox{.}(2019)]%
        {wang2019looking}
\bibfield{author}{\bibinfo{person}{Xueqiang Wang}, \bibinfo{person}{Yuqiong Sun}, \bibinfo{person}{Susanta Nanda}, {and} \bibinfo{person}{XiaoFeng Wang}.} \bibinfo{year}{2019}\natexlab{}.
\newblock \showarticletitle{Looking from the mirror: Evaluating IoT device security through mobile companion apps}. In \bibinfo{booktitle}{\emph{28th USENIX Security Symposium (USENIX Security)}}.
\newblock


\bibitem[Weinmann(2012)]%
        {weinmann2012baseband}
\bibfield{author}{\bibinfo{person}{Ralf-Philipp Weinmann}.} \bibinfo{year}{2012}\natexlab{}.
\newblock \showarticletitle{Baseband Attacks: Remote Exploitation of Memory Corruptions in Cellular Protocol Stacks}. In \bibinfo{booktitle}{\emph{6th USENIX conference on Offensive Technologies (WOOT)}}.
\newblock


\bibitem[Wen et~al\mbox{.}(2020a)]%
        {wen2020plug}
\bibfield{author}{\bibinfo{person}{Haohuang Wen}, \bibinfo{person}{Qi~Alfred Chen}, {and} \bibinfo{person}{Zhiqiang Lin}.} \bibinfo{year}{2020}\natexlab{a}.
\newblock \showarticletitle{Plug-N-Pwned: Comprehensive vulnerability analysis of OBD-II dongles as a new Over-the-Air attack surface in automotive IoT}. In \bibinfo{booktitle}{\emph{29th USENIX Security Symposium (USENIX Security)}}.
\newblock


\bibitem[Wen and Lin(2023)]%
        {wen2023egg}
\bibfield{author}{\bibinfo{person}{Haohuang Wen} {and} \bibinfo{person}{Zhiqiang Lin}.} \bibinfo{year}{2023}\natexlab{}.
\newblock \showarticletitle{Egg hunt in Tesla infotainment: a first look at reverse engineering of Qt binaries}. In \bibinfo{booktitle}{\emph{32nd USENIX Security Symposium (USENIX Security)}}.
\newblock


\bibitem[Wen et~al\mbox{.}(2020b)]%
        {wen2020firmxray}
\bibfield{author}{\bibinfo{person}{Haohuang Wen}, \bibinfo{person}{Zhiqiang Lin}, {and} \bibinfo{person}{Yinqian Zhang}.} \bibinfo{year}{2020}\natexlab{b}.
\newblock \showarticletitle{Firmxray: Detecting bluetooth link layer vulnerabilities from bare-metal firmware}. In \bibinfo{booktitle}{\emph{27th ACM SIGSAC conference on computer and communications security (CCS)}}.
\newblock


\bibitem[Wen et~al\mbox{.}(2023)]%
        {wen2023thwarting}
\bibfield{author}{\bibinfo{person}{Haohuang Wen}, \bibinfo{person}{Phillip Porras}, \bibinfo{person}{Vinod Yegneswaran}, {and} \bibinfo{person}{Zhiqiang Lin}.} \bibinfo{year}{2023}\natexlab{}.
\newblock \showarticletitle{Thwarting Smartphone SMS Attacks at the Radio Interface Layer}. In \bibinfo{booktitle}{\emph{30th Network and Distributed System Security Symposium (NDSS)}}.
\newblock


\bibitem[Wen et~al\mbox{.}(2020c)]%
        {wen2020automated}
\bibfield{author}{\bibinfo{person}{Haohuang Wen}, \bibinfo{person}{Qingchuan Zhao}, \bibinfo{person}{Qi~Alfred Chen}, {and} \bibinfo{person}{Zhiqiang Lin}.} \bibinfo{year}{2020}\natexlab{c}.
\newblock \showarticletitle{Automated cross-platform reverse engineering of CAN bus commands from mobile apps}. In \bibinfo{booktitle}{\emph{27th Network and Distributed System Security Symposium (NDSS)}}.
\newblock


\bibitem[wishihab(2018)]%
        {android_rat}
\bibfield{author}{\bibinfo{person}{wishihab}.} \bibinfo{year}{2018}\natexlab{}.
\newblock \bibinfo{title}{Remote Access Tool Trojan List - Android}.
\newblock \bibinfo{howpublished}{\url{https://github.com/wishihab/Android-RATList}}.
\newblock


\bibitem[Yang and Yang(2012)]%
        {yang2012leakminer}
\bibfield{author}{\bibinfo{person}{Zhemin Yang} {and} \bibinfo{person}{Min Yang}.} \bibinfo{year}{2012}\natexlab{}.
\newblock \showarticletitle{Leakminer: Detect information leakage on android with static taint analysis}. In \bibinfo{booktitle}{\emph{3rd World Congress on Software Engineering (WSSE)}}.
\newblock


\bibitem[Zeng et~al\mbox{.}(2012)]%
        {zeng2012design}
\bibfield{author}{\bibinfo{person}{Yuanyuan Zeng}, \bibinfo{person}{Kang~G Shin}, {and} \bibinfo{person}{Xin Hu}.} \bibinfo{year}{2012}\natexlab{}.
\newblock \showarticletitle{Design of SMS commanded-and-controlled and P2P-structured mobile botnets}. In \bibinfo{booktitle}{\emph{5th ACM Conference on Security and Privacy in Wireless and Mobile Networks (WiSec)}}.
\newblock


\end{thebibliography}
